\documentclass[10pt,journal,compsoc]{IEEEtran}

\ifCLASSOPTIONcompsoc
  \usepackage[nocompress]{cite}
\else
  \usepackage{cite}
\fi

\ifCLASSINFOpdf
   \usepackage[pdftex]{graphicx}
   \graphicspath{{figfinal/}{./}}
   \DeclareGraphicsExtensions{.pdf,.jpeg,.png}
\else
   \usepackage[dvips]{graphicx}
\fi

\usepackage[cmex10]{amsmath}

\usepackage{algorithm}
\usepackage{algorithmic}

\usepackage{array}

\usepackage{verbatim}

\usepackage{bm}
\usepackage{color}

\begin{document}
%
\title{Continuous-Scale Kinetic Fluid Simulation}
%
%
%
%

\author{Wei Li,
	Kai Bai,
      and Xiaopei Liu
\IEEEcompsocitemizethanks{
\IEEEcompsocthanksitem Wei Li, Kai Bai are both with the School
of Information Science and Technology, ShanghaiTech University, and Shanghai Institute of Microsystem and Information Technology, Chinese Academy of Science, Shanghai, China, as well as University of Chinese Academy of Sciences.\protect\\
E-mail: \{liwei, baikai\}@shanghaitech.edu.cn.
\IEEEcompsocthanksitem Xiaopei Liu is with the School
of Information Science and Technology, ShanghaiTech University, Shanghai, China.\protect\\
E-mail: liuxp@shanghaitech.edu.cn and aurorean.xp@gmail.com
\IEEEcompsocthanksitem Corresponding author: Xiaopei Liu.

}}


%
%

\markboth{Accepted by IEEE Transactions on Visualization and Computer Graphics}%
{Li \MakeLowercase{\textit{et al.}}}
%



\IEEEtitleabstractindextext{%
	\begin{abstract}
		Kinetic approaches, i.e., methods based on the lattice Boltzmann equations, have long been recognized as an appealing alternative for solving incompressible Navier-Stokes equations in computational fluid dynamics.
		However, such approaches have not been widely adopted in graphics mainly due to the underlying inaccuracy, instability and inflexibility.
		In this paper, we try to tackle these problems in order to make kinetic approaches practical for graphical applications.
		To achieve more accurate and stable simulations, we propose to employ the non-orthogonal central-moment-relaxation model, where we develop a novel adaptive relaxation method to retain both stability and accuracy in turbulent flows.
		To achieve flexibility, we propose a novel continuous-scale formulation that enables samples at arbitrary resolutions to easily communicate with each other in a more continuous sense and with loose geometrical constraints, which allows efficient and adaptive sample construction to better match the physical scale.
		Such a capability directly leads to an automatic sample construction which generates static and dynamic scales at initialization and during simulation, respectively.
		This effectively makes our method suitable for simulating turbulent flows with arbitrary geometrical boundaries.
		Our simulation results with applications to smoke animations show the benefits of our method, with comparisons for justification and verification.
		
	\end{abstract}
	
	}

\maketitle

\IEEEdisplaynontitleabstractindextext

%
\IEEEpeerreviewmaketitle

\section{Introduction}
\label{sec:intro}

Fluid simulation in graphics has evolved for more than a decade.
Since the pioneering work of \cite{Stam-1999}, fluid simulation methods have been developed significantly, among which directly solving the incompressible Navier-Stokes equations (INSE) can be considered as the standard and very popular approach to simulating fluid flows.
However, under turbulent conditions where the flow usually has small viscosity and thus high Reynolds number, accurately resolving the small-scale turbulence details in INSE effectively without numerical diffusion becomes a great challenge.
In order to tackle this problem, different methods have been proposed in the literature~\cite{Losasso-2004,Selle-2005,Schechter-2008,WeiBmann-2010,Pfaff-2012,Zhang-2015}, but usually at a cost of more algorithmic and computational complexity.
Some of them~\cite{Fedkiw-2001,Bridson-2007,Kim-2008}, which are mainly based on noise models, may not fully respect the underlying physics, making the simulated results unnatural in some circumstances.

While a large number of methods have been proposed to directly solve INSE, there exist other alternatives which can bypass the difficulty of nonlinear advection and global pressure solve in INSE, making the underlying solution simpler and sometimes more accurate.
One of these alternatives is the kinetic approach based on the lattice Boltzmann equations (LBE)~\cite{Chen-1998}.
Researchers from the computational fluid dynamics (CFD) field are interested in such a method, since it transforms INSE into a computationally easier set of linear PDE system with a nonlinear source term, and without any global pressure solve.
In addition, it is usually formulated by an explicit time evolution with a constant time step, which is much simpler to solve with only local updating dynamics.
This facilitates the development of a simple and conservative numerical scheme based on LBE, which is formulated as:
\begin{equation}
f_i(\mathbf{x}+\mathbf{c}_i\Delta t , t+\Delta t) -f_i(\mathbf{x},t) =\Omega_i(\rho,\mathbf{u}) , \label{eq:lbe}
\end{equation} 
where $f_i$ is the $i$-th velocity distribution function associated with the $i$-th lattice velocity $\mathbf{c}_i$; $\Omega_i$ is the collision operator designed to be conservative and important to approximate INSE; $\Delta x=\Delta t$ with proper $\mathbf{c}_i$ should be strictly satisfied for stability (CFL=1, with CFL a number defined as CFL=$u\Delta t/\Delta x$~\cite{Courant-1967}, where $u$ is the flow speed; larger CFL number indicates larger time stepping and faster simulation over time); $\rho$ and $\mathbf{u}$ are the macroscopic density and velocity computed as $\rho=\sum_i f_i$ and $\mathbf{u}=\sum_i \mathbf{c}_i f_i / \rho$.
Eq.~\ref{eq:lbe} is usually accurate and stable in the incompressible limit (the entire flow speed is small, e.g., $0.1$, as compared to the speed of sound $c_s = 1/\sqrt{3}$ in normalized lattice units).

Solving fluid flows using LBE has several known advantages.
In addition to the benefits introduced before, the local treatment of boundary conditions leads to effective simulations with arbitrary geometrical boundaries and with efficient implementations for parallelism.
It is well known that the stability and accuracy of Eq.~\ref{eq:lbe} largely depends on the modeling of $\Omega_i$, where traditionally the Bhatnagar-Gross-Krook (BGK)~\cite{Chen-1998} and multiple-relaxation-time (MRT)~\cite{DHumieres-2002} models were often used, but the inherent inaccuracy and instability prevent their wide-spread use in graphics.
Thuerey et al.~\cite{Thuerey-2006,Thuerey-2008} and Zhao et al.~\cite{wei-2004,Zhao-2007} ever promoted the use of LBE in graphics, but only for flows with moderate Reynolds numbers.
Liu et al.~\cite{Liu_TVCG_2014} and Guo et al.~\cite{Guo_TVCG_2017} also proposed to use LBE for graphical applications, but with more empirical approaches, resulting in less realistic results in some circumstances.

In this paper, we aim to solve fluid flows using the kinetic approach. 
However, to promote the practical use of such an approach for graphical fluid flow simulations, we need to improve the stability, accuracy and flexibility.
To achieve higher stability and accuracy, a more suitable collision operator $\Omega_i$ should be developed in order to approximate INSE more closely and suppress ghost and ringing artifacts more appropriately.
To achieve computational flexibility, easy and more continuous adaptation of sample resolutions should be achieved in order to resolve the physical details during simulations.
The solution to the above problems form the core contributions of this paper:
\begin{itemize}
	\item
	{
		To significantly increase the stability and accuracy over the traditional BGK and MRT models, we employ a novel non-orthogonal central-moment-relaxation (CMR) model for LBE~\cite{Rosis-2017}, which has higher convergence to INSE \textit{given appropriate relaxation parameters}, and with simpler algebra.
		However, like in MRT model, not all the relaxation parameters can give satisfactory results for CMR model, and determining the appropriate relaxation parameters is \textit{crucial} to retain turbulence details with both stability and accuracy.
		With the observation that CMR model conducts relaxations independently for different orders of moments,
		and high-order moments influence the turblence details significantly, we propose to determine the high-order relaxation parameters adaptively according to the local velocity gradient in order to produce stable yet accurate flows under turbulent conditions, which makes the method practical for graphical applications.
	}
	
	\vspace{0.2cm}
	\item
	{
		On the other hand, traditional LBE simulations usually solve the fluid flows with a uniform lattice, which is hard to adapt computations to spatially and temporally varying physical details.
		Multi-block approaches~\cite{Peng-2006,Thuerey-2008} solve this problem by dividing a large-scale cell into a multiple of \textit{integer} small-scale cells with strict geometrical alignment along the scale boundary, which is not flexible and violates the \textit{continuous-scale} nature of fluid flows, resulting in undesirable discontinuous structures in turbulent flows.
		In this work, we propose a novel \textit{continuous-scale} formulation, which allows samples in \textit{arbitrary} scales to communicate with each other without strict spatial constraint, where mappings and interpolations of distribution functions are properly handled.
		By ``continuous-scale'', we mean that the ratio between different scales could be arbitrary rather than only integral.
		This immediately enables a fully automatic scheme to first generate static scales at initialization that transit more continuously based on domain geometries and inlet positions, and then refine the scales dynamically at runtime during the simulation, which is particularly useful in boundary-induced turbulent flow simulations.		
	}
\end{itemize}

\begin{figure}[t]
	\centering
	\includegraphics[width=\columnwidth]{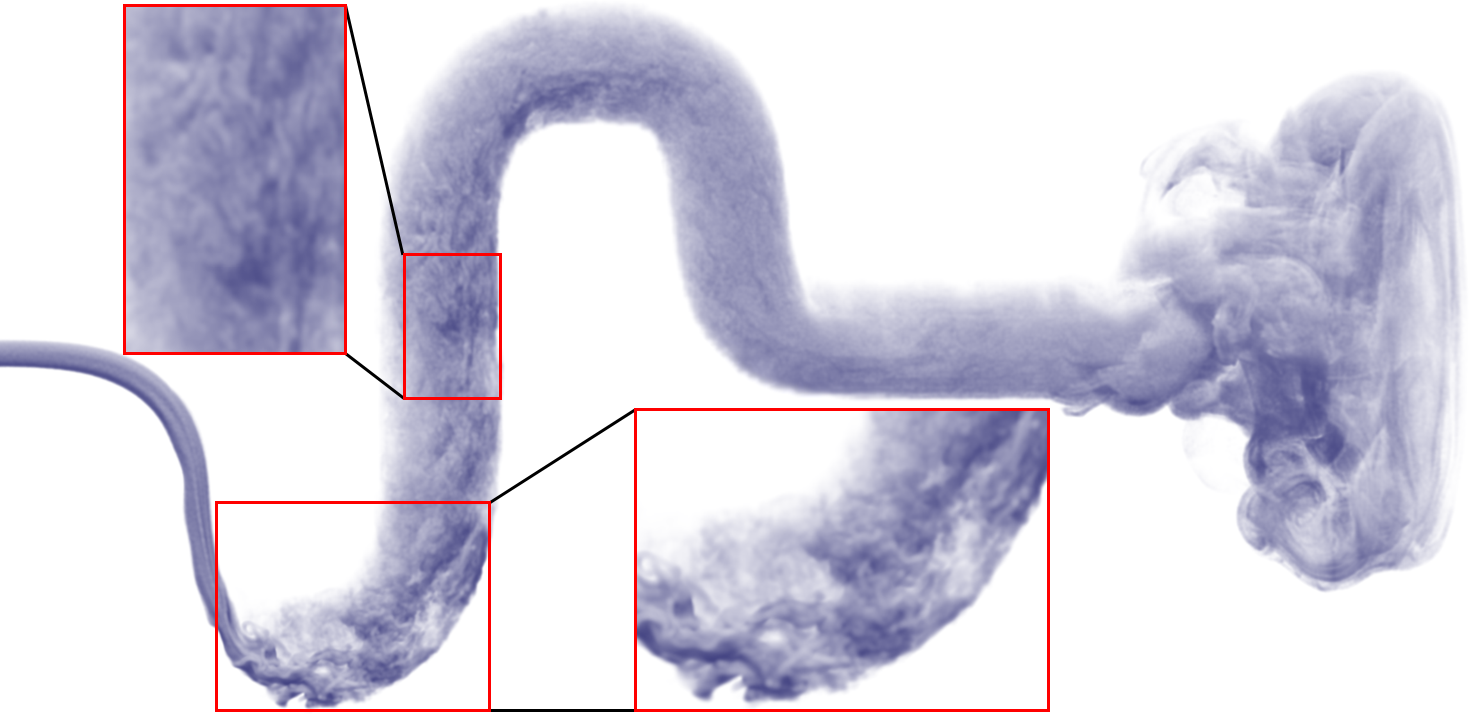}
	\caption{Smoke simulation using our continuous-scale kinetic fluid simulator. The smoke is injected from the inlet of the tube and follows the air flow into the tube. The boundary of the tube generates chaotic smoke turbulence patterns. Note that we can easily place higher resolution samples along the tube boundary to have more accurate computations. This example also demonstrates the capability of our method to simulate flows under complex geometric boundaries.
	}
	
	\label{fig:teaser}
\end{figure}

To justify our arguments, we present the results by applying our method to smoke simulations, where stable solutions can be readily achieved with sufficient turbulence details and over arbitrary geometries, see Fig.~\ref{fig:teaser} for an example of our simulated smoke passing through a turning tube, where the boundary layer induces small vortices, and higher resolution samples are placed along the tube boundary.
Note that in our approach, no turbulence nor noise models are used to resolve the fluid details.
As a verification, our results are compared to the existing methods for smoke simulations~\cite{Selle-2008,Zhang-2015,Zhang-2016}, which suggest that more appropriate visual details can be achieved with less number of samples and with higher computational efficiency.

\section{Related Work}
\label{sec:related_work}

There are a vast number of research work on fluid simulations in graphics.
Here, we summarize the works that are quite related to our work on single-phase turbulent flow simulations, and ignore those that deal with multi-phase flows.
In addition, we categorize the related methods into direct approaches where INSE is directly solved, and indirect approaches where other equations are solved in order to approximate the solution of INSE.

\subsection{Direct approach}
\label{sec:related_work_direct_approach}

To solve INSE directly, Stam~\cite{Stam-1999} proposed the unconditionally stable semi-Lagrangian advection scheme, with the major drawback of excessive numerical diffusion.
To overcome such a problem, many different algorithms have been proposed.
Vorticity confinement~\cite{John-1994,Fedkiw-2001} was the early attempt to add fluid details by artificial force, which was later extended by noise-based approaches~\cite{Bridson-2007,Kim-2008} where the Kolmogorov energy spectrum is respected.
To have better results around object boundary, turbulence models~\cite{Schechter-2008,Pfaff-2010} and pre-computed artificial boundary layer method~\cite{Pfaff-2009} were proposed.
However, all of these methods do not fully respect the underlying physics, making the simulation sometimes not realistic.

To preserve fluid details without artificial modeling, different classes of methods were proposed.
One class of such methods try to increase the accuracy of non-linear advection where BFECC~\cite{Kim-2005} , MacCormack~\cite{Selle-2008} and high-order WENO schemes~\cite{Wang-2008} , as well as improved high-order constrained interpolation profile (CIP) methods~\cite{Song-2005,Kim-2008} were used.
Heo and Ko~\cite{Heo-2010} combined polynomial representation with a high-order re-initialization method to preserve detailed structures of the fluid interface.
However, the more widely-adopted approach is the hybrid method where advection is solved using particles while pressure and other parts are solved using grids~\cite{Zhu-2005,Raveendran-2011}.
Jiang et al.~\cite{Jiang-2015} presented a novel technique to preserve linear and angular momentum in the hybrid approach in order to better resolve the details.
Such a technique has later been extended to a more generalized local function to greatly improve the energy and vorticity conservation~\cite{Fu-2017}.
Vortex methods are also appealing to preserve fluid details~\cite{Park-2005,Golas-2012}.
Vortex filaments~\cite{WeiBmann-2010} and vortex sheets~\cite{Brochu-2012,Pfaff-2012} are both effective ways to solve for turbulent flows with reduced computation.
To improve the efficiency for solving Poisson equation, Zhang and Bridson~\cite{ZhangXX-2014} proposed a hybrid PPPM algorithm.
More recently, Zhang et al.~\cite{Zhang-2015} proposed the IVOCK scheme to preserve more turbulence details based on the velocity correction from the vorticity equation.

In addition to the above methods, adaptive approaches~\cite{Losasso-2004,Mullen-2009,Lentine-2010} try to put more computations on fine structures to capture flow details.
Zhu et al.~\cite{Zhu-2013} presented an adaptive grid to create a far-field coarse grid with fine grid at the focus of the simulation.
Setaluri et al.~\cite{Setaluri-2014} introduced a new data structure for adaptive grids with compact storage and efficient stream processing .
Recently, Zhang et al.~\cite{Zhang-2016} proposed an adaptive particle-grid scheme to capture boundary layer dynamics more accurately.

There are also many Lagrangian particle solvers for INSE, which are mainly based on smoothed particle hydrodynamics (SPH), and are naturally spatially adaptive, e.g., 
Becker~\cite{Becker-2007} presented a weakly compressible form of the SPH method for fluid flows based on the Tait equation; Solenthaler~\cite{Solenthaler-2009} presented a novel incompressible SPH method for fluid simulations based on prediction-correction scheme;
Ihmsen~\cite{Ihmsen2-2014} proposed a novel formulation of the projection method for SPH; and Winchenbach~\cite{Winchenbach-2017} introduced a novel method for adaptive incompressible SPH simulations.

Another new approach to solving INSE is the data-driven approach based on machine learning~\cite{Jeong-2015,Chu-2017}, which can produce fast solutions, but the results may contain unexpected artifacts.

Compared to the existing approaches above, our method does not solve the INSE directly, which overcomes the difficulty of handling nonlinear advection and global pressure solve with sufficient accuracy and stability.
More importantly, we propose an efficient adaptive scale refinement formulation, which allows continuous-scale construction with loose geometrical constraint. This facilitates flexible and more efficient simulations.

\subsection{Indirect approach}
\label{sec:related_work_indirect_approach}

INSE can also be solved indirectly by other model equations, among which the kinetic approach based on LBE is one alternative.
The early work of LBE-based approach was pioneered by BGK model~\cite{Chen-1998}.
To improve stability and accuracy, MRT model~\cite{DHumieres-2002} was proposed.
However, the most significant progress for LBE modeling is the cascaded model with central-moment relaxation~\cite{Geier-2006,Daniel-2014}.
Very recently, De Rosis~\cite{Rosis-2017} proposed a non-orthogonal central-moment-relaxation model with simple algebra.
Turbulence models, on the other hand, can also be used to stabilize LBE and retain fluid details especially in coarse grid simulations~\cite{Liu-2012,Geller-2013}, but may introduce numerical artifacts.
In this paper, we employ the non-orthogonal central-moment relaxation model, but propose an adaptive relaxation scheme without the aid of turbulence models for graphical flow simulations, which respects the underlying physics more appropriately.

To enable adaptive computation, multi-block-based grid refinement~\cite{Filippova-1998,Dupuis-2003,Peng-2006,Zhao-2007,Thurey-2009}, and unstructured mesh formulations~\cite{Fan-2005,Qu-2010} were proposed for LBE.
While multi-block formulation lacks scale-continuity and has strict alignment constraint between scales, unstructured mesh formulation requires complicated meshing/re-meshing process, which is difficult for dynamic refinement.
Compared to these methods, we propose a novel continuous-scale formulation that allows arbitrary scales to communicate with each other without strict spatial constraint.
As described in Section~\ref{sec:intro}, this immediately allows flexible scale construction and dynamic refinement to resolve turbulent flow details more appropriately with structure continuity.

There are some other works that deal with the interaction between fluid flows and solid objects using LBE approach in graphics.
For example, Wei et al.~\cite{wei-2004} presented an approach for simulating natural dynamics that emerge from the interaction between a flow field and immersed objects.
Zhao et al.~\cite{Zhao-2007} provided a physically-based framework for simulating natural phenomena related to heat interaction between objects and the surrounding air.
In addition to kinetic approaches, incompressible nonlinear Schr\"{o}dinger equation has been recently employed in graphics to solve for inviscid fluid flows~\cite{Chern-2016} with more accurate advection.

\section{Fundamentals}

Before presenting our approach, we first introduce the fundamentals on the non-orthogonal central-moment relaxation model for LBE as well as the multi-block formulations.
They also serve as a reference to differentiate our formulation with the existing ones.

\subsection{Non-orthogonal central-moment relaxation model}
\label{sec:central_moment_relax}

As introduced before, central-moment relaxation (CMR) models have superior performance than BGK and MRT models in terms of stability and accuracy, and in particular, we employ the non-orthogonal CMR model~\cite{Rosis-2017}, which is constructed with simpler algebra and sufficient stability.

\begin{figure}[t]
	\centering
	\includegraphics[width=0.55\columnwidth]{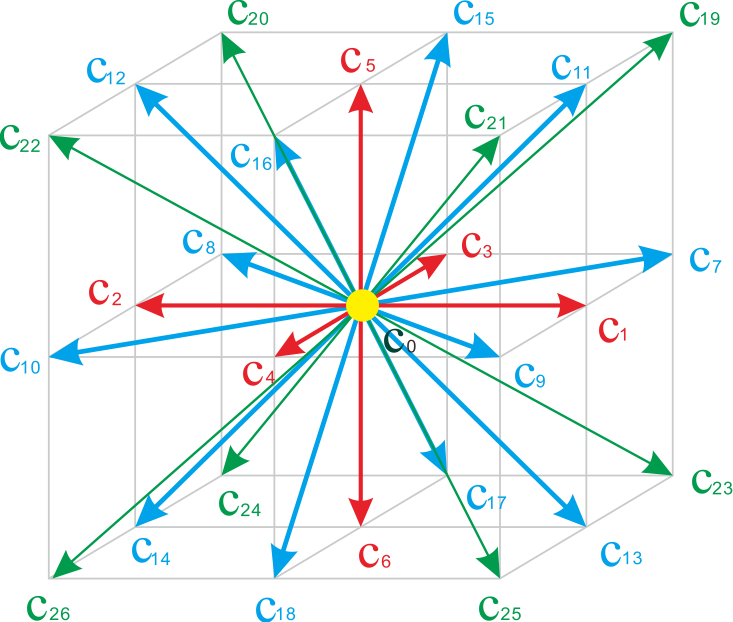}
	\caption{The lattice structure (D3Q27) used in our 3D kinetic fluid simulations, where $\mathbf{c}_i$ is the discretized microscopic velocity. Note that each $f_i$ is associated with a corresponding $\mathbf{c}_i$.}
	\label{fig:lattice_structure}
\end{figure}

Unlike the traditional MRT model, the non-orthogonal CMR model constructs the central-moment space with translated lattice velocities: $\bar{\mathbf{c}}_i=\mathbf{c}_i-\mathbf{u}$, where $\mathbf{c}_i$ is the original lattice velocity (we use D3Q27 lattice velocity model for 3D simulations, see Fig.~\ref{fig:lattice_structure}), and $\mathbf{u}$ is the macroscopic velocity.
The collision is performed in the central-moment space, and thus a transformation of the distribution function $f_i$ into central-moment space should be performed.

To obtain such a transformation, a matrix $\mathbf{M}$ is first constructed as: $\mathbf{M}_{i,j}=\bar{\mathbf{c}}^m_{x,i} \bar{\mathbf{c}}^n_{y,i} \bar{\mathbf{c}}^p_{z,i}$, where $\{x,y,z\}$ indexes the corresponding velocity component; $m,n,p \in \{0,1,2\}$ are the orders of moments, and $j=(m+1)(n+1)(p+1)-1$ indexes different components.
Then, we perform $\mathbf{m}=\mathbf{M}^T \mathbf{f}$ where $\mathbf{f}$ and $\mathbf{m}$ aggregate all values for $f_i$ and $m_i$ to perform the transformation.
By constructing an inverse matrix $\mathbf{T}=(\mathbf{M}^T)^{-1}$, we can transform the moment vector back to the distribution functions by $\mathbf{f}=\mathbf{T}\mathbf{m}$.
Note that the analytical forms for both $\mathbf{M}$ and $\mathbf{T}$ can be explicitly obtained, which are directly used during the simulation.

To model collision, the equilibrium state is constructed in central-moment space as $\mathbf{m}^{eq}$ and the collision vector $\mathbf{\Omega}$, which contains all the collision operators $\Omega_i$, is constructed by a relaxation process in central-moment space as:
\begin{equation}
\mathbf{\Omega}=-\mathbf{T}\mathbf{S}(\mathbf{m}-\mathbf{m}^{eq}), \label{eq:cmr_collision}
\end{equation}
where $\mathbf{S}$ is a diagonal relaxation matrix.
Note that some relaxation parameters are related to the kinematic viscosity:
\begin{equation}
\nu =\frac{1}{3} \left(\frac{1}{\mathbf{S}_i} - \frac{1}{2}\right), \,\,\,\,   i\in\{4,5,6,7,8\},
\label{eq:viscosity}
\end{equation}
thus, $\mathbf{S}_i = S (i\in\{4,5,6,7,8\})$ are all of the same value.
Parameters for conservative quantities $\mathbf{S}_i (i\in\{0,1,2,3\})$ can be arbitrary, and we set them to be 0.
Other parameters $\mathbf{S}_i (i>8)$ are for high-order moments, which can be freely tuned within the range $(0,2)$ to achieve different accuracy and stability.
The specific forms of $\mathbf{M}$, $\mathbf{T}$, $\mathbf{m}$ and $\mathbf{m}^{eq}$ can be found in Appendix.

\subsection{Multi-block lattice Boltzmann formulation}
\label{sec:multi_block_formulation}

\begin{figure}[t]
	\centering
	\includegraphics[width=\columnwidth]{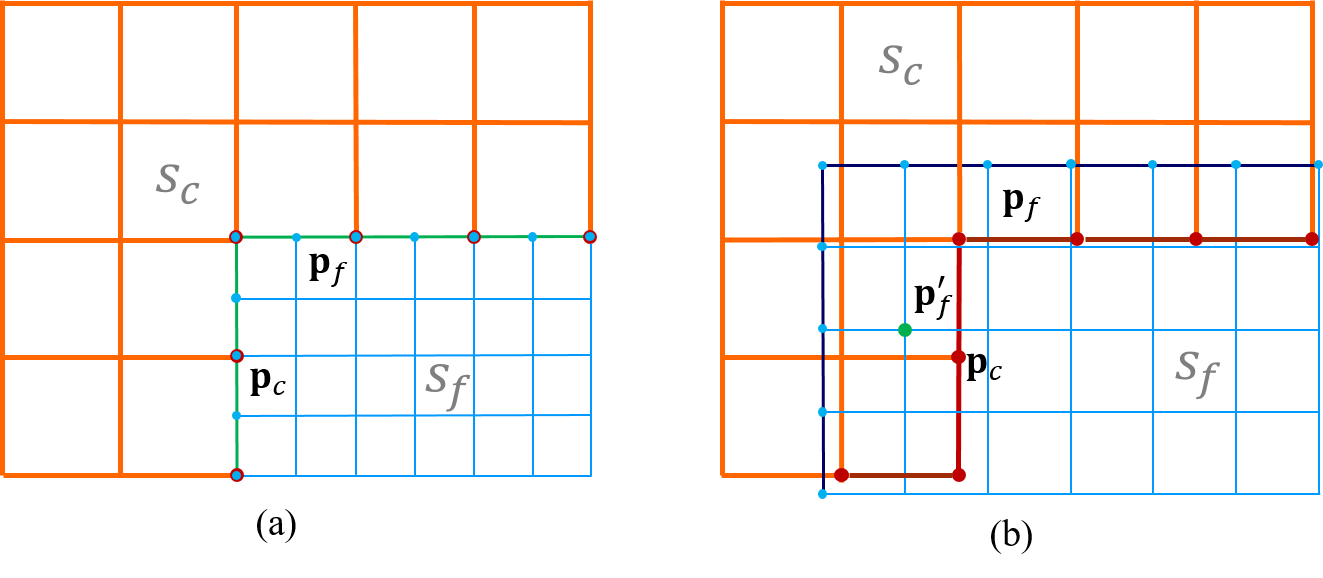}
	\caption{Spatial scale mapping for multi-block and our continuous-scale simulations: (a) multi-block formulation: fine and coarse scales coincide with their boundaries (the green line) and $\alpha$ is strictly an integer value ($\alpha=2$ in this example). (b) our continuous-scale formulation: $\alpha$ can be arbitrary ($\alpha=1.4$ in this example), and the scales do not need to coincide with their boundaries; in this case, interpolation from the nearby samples is required before mapping.}
	\label{fig:multiblock-continuous-scale}
\end{figure}

As mentioned in Section~\ref{sec:related_work_indirect_approach}, adaptive approaches have been proposed to simulate LBE in order to save computation, among which multi-block formulations were often used due to simplicity especially considering dynamic refinement.
In multi-block formulation, the uniform grid is subdivided in an octree manner.
Fig.~\ref{fig:multiblock-continuous-scale} (a) gives an example of grid refinement with two scales, where the coarse scale (orange one) is subdivided into the fine scale (blue one), with their boundaries matched (see the green line), and the ratio between different scales be strictly integral.

The idea behind multi-block formulation is to keep local Reynolds number invariant at the same sample point between different scales, which results in a mapping of $f_i$ at overlapped sample points between different scales (the green line in Fig.~\ref{fig:multiblock-continuous-scale} (a)).
There have been derivations of such a mapping for BGK~\cite{Yu-2002,Filippova-1998} and MRT~\cite{Peng-2006, Geller-2013} models, but no derivation for non-orthogonal CMR models yet.

Our continuous-scale approach is based on multi-block formulation.
To explain it, we first define some notations:
$s$ indicates a specific scale; $c$ and $f$ indicate the coarse and fine scales; $\Delta x_s$ and $\Delta t_s$ denote the grid spacing and time step at scale $s$; $\alpha$ is the ratio of spacings between two scales.

To obtain the mapping of $f_i$ for MRT model, the invariance of local Reynolds numbers between coarse and fine scales leads to~\cite{Peng-2006}:
\begin{equation}
1/\mathbf{S}_i^f-1/2= \alpha \left(1/\mathbf{S}_i^c-1/2\right)\label{eq:S_relation},
\end{equation}
with $\alpha=\Delta x_c/\Delta x_f$.
To further maintain continuity of macroscopic variables across scales, it is derived in \cite{Peng-2006} that:
\begin{equation}
\mathbf{m}_i^{neq,c} = \alpha \frac{\mathbf{S}_i^f}{\mathbf{S}_i^c} \mathbf{m}_i^{neq,f}, \label{eq:m_eq_relation}
\end{equation}
which results in a diagonal mapping matrix of all the non-equilibrium components in moment space from fine to coarse scales as: $\mathbf{m}^{neq,c}=\mathbf{K}^{f \rightarrow c}\mathbf{m}^{neq,f}$.
Similarly, we can obtain a diagonal mapping matrix from coarse to fine scales.
Returning back to the space of $\mathbf{f}$ and following \cite{Peng-2006}, the mapping from fine to coarse scales after collision is:
\begin{equation}
\begin{aligned}
\tilde{\mathbf{f}}^c = &\mathbf{T}\mathbf{m}^{c} - \mathbf{T}\mathbf{S}^c(\mathbf{m}^c - \mathbf{m}^{eq,c})\\& 	
= \mathbf{T}(\mathbf{m}^{eq,c} + \mathbf{m}^{neq,c}) - \mathbf{T}\mathbf{S}^c\mathbf{m}^{neq,c}\\&
= \mathbf{T}\mathbf{m}^{eq,f} + \mathbf{T}(\mathbf{I} - \mathbf{S}^c)\mathbf{K}^{f \rightarrow c}\mathbf{m}^{neq,f} ,
\end{aligned}
\end{equation}
where tilde indicates post-collision state and $\mathbf{I}$ is the identity matrix.
This immediately expresses $\mathbf{m}^{neq,c}$ and $\mathbf{m}^{neq,f}$ as:	
\begin{equation}
\begin{aligned}
\mathbf{m}^{neq,c}&=(\mathbf{I} - \mathbf{S}^c)^{-1}(\tilde{\mathbf{m}}^{c}-\mathbf{m}^{eq,c}) , \\
\mathbf{m}^{neq,f}&=(\mathbf{I} - \mathbf{S}^f)^{-1}(\tilde{\mathbf{m}}^{f}-\mathbf{m}^{eq,f}) .
\end{aligned}
\end{equation}
Then, we can rewrite $\tilde{\mathbf{f}}^c$ as:	
\begin{equation}
\begin{aligned}
\tilde{\mathbf{f}}^c = \mathbf{T}\left(\mathbf{m}^{eq,f}+(\mathbf{I}-\mathbf{S}^c)\mathbf{K}^{f \rightarrow c}\frac{\tilde{\mathbf{m}}^{f}-\mathbf{m}^{eq,f}}{\mathbf{I} - \mathbf{S}^f}\right) .
\end{aligned}
\end{equation}
Defining $\hat{\mathbf{K}}^{f \rightarrow c} =  (\mathbf{I}-\mathbf{S}^c)\mathbf{K}^{f \rightarrow c}(\mathbf{I} - \mathbf{S}^f)^{-1}$	and $\tilde{\mathbf{m}}^{neq,f}=\tilde{\mathbf{m}}^{f}-\mathbf{m}^{eq,f}$,
the mapping from fine to coarse scales can be rewritten as:
\begin{equation}
\begin{aligned}
\tilde{\mathbf{f}}^c=\mathbf{T}\left(\mathbf{m}^{eq,f}+\hat{\mathbf{K}}^{f \rightarrow c}\tilde{\mathbf{m}}^{neq,f}\right) .
\end{aligned}
\end{equation}
The mapping from coarse to fine scales can be constructed and formulated similarly.

\begin{figure*}[t]
	\centering
	\includegraphics[width=0.95\textwidth]{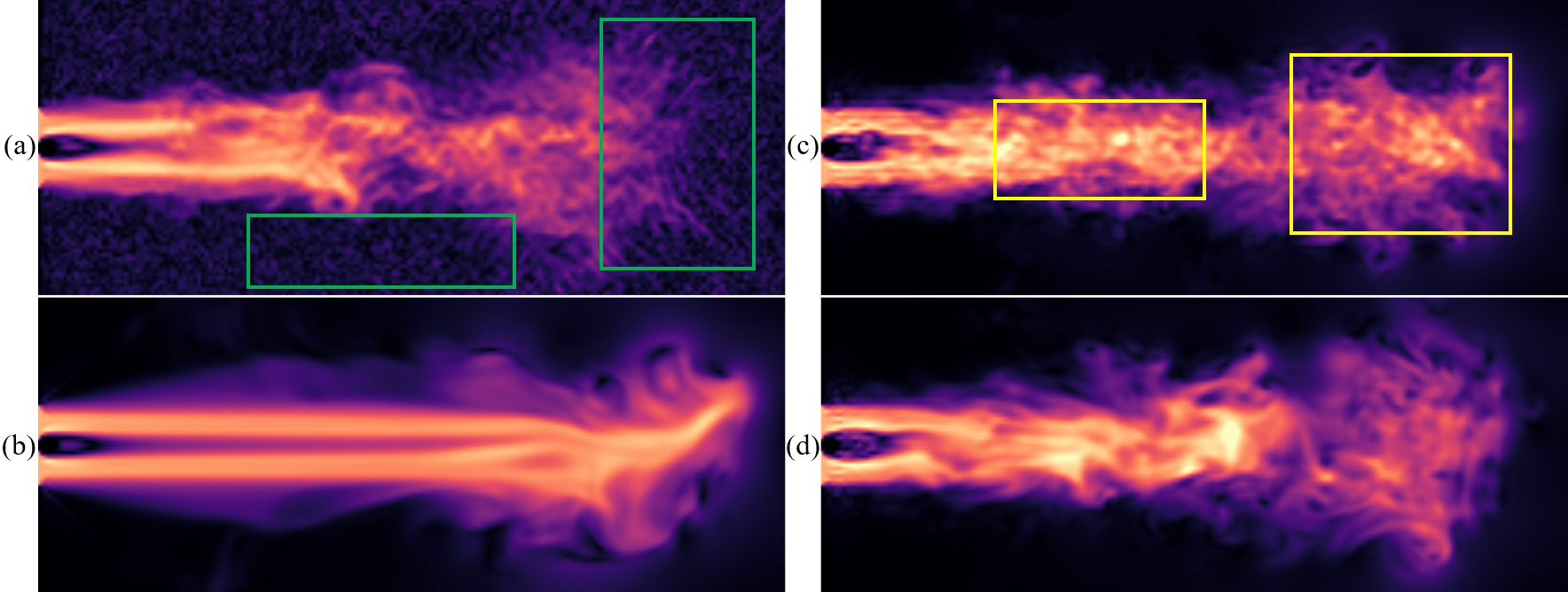}
	\caption{Comparison of cross-section 3D velocity fields among simulations from MRT and non-orthogonal CMR models with different relaxation parameter settings (resolution: $196 \times 98 \times 98$, viscosity: $\nu=10^{-4}$). (a) MRT with Smagorinsky model where ringing artifacts are strong and all over the field (see the green boxes); (b) non-orthogonal CMR model with the original relaxation parameter setting from \cite{Rosis-2017}, which over-smooths the flow field; (c) non-orthogonal CMR model with our fixed relaxation parameter setting, where ringing artifacts may still persist (see the yellow boxes); (d) non-orthogonal CMR model with our adaptive relaxation, which suppresses most ringing artifacts while preserving turbulence details.}
	\label{fig:mrt_cmr_compare}
\end{figure*}

To summarize, in general, given two scales $s_i$ and $s_j$, the mapping from $s_i$ to $s_j$ after collision can be given by:
\begin{equation}
\tilde{\mathbf{f}}^{s_j} = \mathbf{T}\left(\mathbf{m}^{eq,s_i}+\hat{\mathbf{K}}^{s_i \rightarrow s_j}\tilde{\mathbf{m}}^{neq,s_i}\right) , \label{eq:s1_to_s2_mapping}
\end{equation}
where 
\begin{equation}
\hat{\mathbf{K}}^{s_i \rightarrow s_j} =  (\mathbf{I}-\mathbf{S}^{s_j})\mathbf{K}^{s_i \rightarrow s_j}(\mathbf{I}-\mathbf{S}^{s_i})^{-1},
\end{equation}
and $\mathbf{K}^{s_i \rightarrow s_j}$ is a diagonal matrix with
\begin{equation}
diag\{\mathbf{K}^{s_i \rightarrow s_j}\} =\{1, ..., \frac{\mathbf{S}_{4}^{s_i}}{\alpha\mathbf{S}_{4}^{s_j}}, ... , \frac{\mathbf{S}_{8}^{s_i}}{\alpha\mathbf{S}_{8}^{s_j}}, \\..., 1\},
\end{equation}
and $\alpha=\Delta x_{s_i}/\Delta x_{s_j}$ is the ratio of spacings between scales $s_i$ and $s_j$.

\section{Our Formulations}
\label{sec:contineous-scale-formulation}

Based on the above fundamental descriptions, we derive our own formulation for LBE, which can achieve stable and accurate simulations, with flexible sample placement and refinement, making our simulator more efficient to produce turbulent flows.

\subsection{Adaptive relaxation}
Traditional MRT model for LBE is unable to simulate fluid flows with small viscosity, which is mainly due to the violation of Galilean invariance, leading to ghost modes that induce instability.
Even though turbulence models (e.g., Smagorinsky model) can stabilize the dynamics, it produces strong ringing artifacts, see Fig.~\ref{fig:mrt_cmr_compare} (a), which contaminates the whole velocity field (\textit{readers are suggested to see the supplementary video for more obvious ringing artifacts}).
The non-orthogonal CMR model can significantly reduce the ghost modes, and thus the ringing artifacts, but the selection of the relaxation parameters is \textit{crucial}.
Improper selection of these parameters may lead to over-smoothed results, see Fig.~\ref{fig:mrt_cmr_compare} (b) with the original parameter setting from~\cite{Rosis-2017}.

To preserve turbulence details while reducing ringing artifacts, it is essential that the high-order relaxation parameters $\mathbf{S}_i (i>8)$ should be carefully tuned~\cite{Rosis-2017}.
However, how these parameters are set to maintain stability while preserving turbulence details is still unknown.
By our analysis, we noticed that each high-order relaxation parameter $\mathbf{S}_i (i>8)$ effectively corresponds to a diffusion viscosity $\nu_i'$ similar to Eq.~\ref{eq:viscosity}, which acts like an artificial viscosity to control high order oscillation modes (ringing artifacts).
In practice and with our numerical experiments, to ensure stability and retain accuracy, $\nu_i'$ should be progressively increased with respect to the order, and should be relatively large (e.g., $\nu_{max}'=0.01$) for the highest order parameter $\mathbf{S}_i (i=26)$, and relatively small (e.g., $\nu_{min}'=0.005$) for the lowest order parameters $\mathbf{S}_i (i=9,10,...,16)$.
For orders in between, we linearly interpolate them based on these two values and their corresponding orders.
This can effectively stabilize the dynamics while retaining sufficient turbulence details, but may not be able to fully suppress the ringing, see Fig.~\ref{fig:mrt_cmr_compare} (c) for an example. 

To further reduce ringing artifacts, we propose to perform \textit{adaptive relaxation}, meaning that instead of using fixed relaxation parameters, we adjust the relaxation for high-order moment adaptively according to the flow.
In principle, fluctuating regions may generate strong numerical waves that propagate to other smoother regions, resulting in noticeable ringing artifacts.
To suppress these rings, we can give more diffusion to the smoother regions, which has the effect of preventing rings from propagating out.
Thus, we should set larger $\nu_i'$ for smoother regions, but the inter-relationship among the high-order parameters should be maintained in proportion to our previous setting in order to have stable simulations.
Hence, we uniformly scale the previous artificial viscosity setting $\nu_i' (i>8)$ for high-order moments according to the velocity gradient, and for samples with smaller gradients, larger scaling factors are given in front of the original $\nu_i'$.
This results in the following formulation of the new artificial viscosity $\hat{\nu}'_i$ for $\mathbf{S}_i (i>8)$:
\begin{eqnarray}
\hat{\nu}'_i=\left(a\frac{|\nabla \mathbf{u}|}{g_{max}}+b\right)\nu_i', \,\,\,\, i>8,
\end{eqnarray} 
where {$a=-4$ and $b=5$} are model parameters that can be tuned; $g_{max}\in[0.1,0.13]$ is the maximum gradient magnitude for normalization.

Fig.~\ref{fig:mrt_cmr_compare} (d) shows the simulation result for the velocity field with such an adaptive relaxation, which is clear that ringing artifacts have been significantly suppressed.
Note that such an adaptive relaxation method cannot be applied to the traditional MRT model since the moment of different orders are coupled for MRT, while in non-orthogonal CMR model, these high-order moments are more independent.

\subsection{Continuous-scale formulation}
By using non-orthogonal CMR model with our adaptive relaxation, we can obtain stable turbulent flow simulations with sufficient fine details.
However, such a simulation is only performed on a uniform grid, and as argued before, it is difficult to adapt computations to spatially and temporally varying fluid flows with different physical details.
In reality, a fluid flow may contain both laminar and turbulent regions, as well as the transition region between them.
The variations of flow quantities (such as velocity) are different,
leading to the concept of \textit{``fluid scale''}, which indicates the frequency of such variations.
It is well known that in real fluid flows, the scale variations are continuous~\cite{Pope-2001}.

In principle, the sample resolution should vary with respect to the fluid scale, where turbulent regions should have more samples.
As introduced in Section~\ref{sec:multi_block_formulation}, multi-block formulation~\cite{Peng-2006,Ubertini-2008} has been proposed to achieve this goal.
While such a formulation is relatively simple, the scales do not respect the \textit{continuous-scale nature} of fluid flows.
To retain scale continuity, unstructured-mesh formulations~\cite{Qu-2010} were proposed, but they all inherited the difficulties for mesh construction and adaptive refinement.

In this paper, we propose a novel method from the idea of multi-block formulation, but allow sample resolutions (scales) to be constructed more continuously, with the ratio between different scales $\alpha$ no longer restricted to integer values.
By breaking such a restriction, we have two important benefits: i. the sample scales can be more continuous in order to better respect the physical scale; ii. as depicted in Section~\ref{sec:scale_construct_tracking}, efficient and flexible scale construction and refinement schemes can be developed in order to dynamically adapt sample scales and place more computations on turbulent fine-scale regions.


\subsubsection{Mapping distribution functions}
To achieve continuous-scale formulation, we need first derive the mapping of distribution functions between different scales for the non-orthogonal CMR model with our adaptive relaxation.
Similar as the derivation for MRT model in Section~\ref{sec:multi_block_formulation}, we start from the invariance of local Reynolds numbers between a coarse and a fine scale, which leads to Eq.~\ref{eq:S_relation}.
Then, we will derive the relationship of non-equilibrium states between $\mathbf{m}_i^{neq,c}$ and $\mathbf{m}_i^{neq,f}$ for the non-orthogonal CMR model in order to obtain the mapping, like the relation in Eq.\ref{eq:m_eq_relation}.

From~\cite{Latt-2007}, we know that $f_i = f_i^{eq}+\epsilon f_i^{(1)}+O(\epsilon^2)$, where $\epsilon \ll 1$ can be identified by the Knudsen number~\cite{Kerson-1987}.
We also know that $\epsilon f_i^{(1)}$ is given by:
\begin{equation}
\epsilon f_i^{(1)} = \frac{w_i}{2c_s^4}\mathbf{Q}_i:\mathbf{\Pi}^{(1)}, \label{eq:epsilon_f1}
\end{equation}
where $w_i$ is the lattice weight originally present in BGK model; $c_s$ is the speed of sound, $\mathbf{Q}_i = \mathbf{c}_i \mathbf{c}_i -c_s^2\mathbf{I} $  and $\mathbf{\Pi}^{(1)}= \sum_i \mathbf{c}_i \mathbf{c}_j \epsilon f_i^{(1)}$ are related to the strain rate tensor $\hat{\mathbf{S}}$ through the relation:
\begin{equation}
\mathbf{\Pi}^{(1)} = -2c_s^2\rho\hat{\mathbf{S}}/S, \label{eq:pi_1}
\end{equation}
where $S$ is related to the kinematic viscosity as in Eq.~\ref{eq:viscosity}, and the strain rate tensor is defined as $\hat{\mathbf{S}} = (\nabla \mathbf{u} + (\nabla \mathbf{u})^T)/2$.
Since $f_i \approx f_i^{eq}+\epsilon f_i^{(1)}$,
we can find that $f_i^{neq} = \epsilon f_i^{(1)}$ is proportional to the gradient of the macroscopic velocity, and it is therefore necessary to be rescaled when communicating between different scales.
By assuming $f_{i}^{neq,f} = \lambda f_{i}^{neq,c}$ and using Eqs.~\ref{eq:epsilon_f1} and~\ref{eq:pi_1}, we have:
\begin{equation}
\frac{1}{\mathbf{S}_f}\mathbf{Q}_i:\hat{\mathbf{S}}_f = \lambda \frac{1}{\mathbf{S}_c}\mathbf{Q}_i:\hat{\mathbf{S}}_c,
\end{equation}
where $\hat{\mathbf{S}}_f$ and $\hat{\mathbf{S}}_c$ represent the same strain rate tensor with lattice units at fine and coarse scales respectively, which can be renormalized by $\Delta x_f$ and $\Delta x_c$, leading to:
\begin{equation}
\frac{\Delta x_f}{\mathbf{S}_f}\mathbf{Q}_i:\hat{\mathbf{S}} = \lambda \frac{\Delta x_c}{\mathbf{S}_c}\mathbf{Q}_i:\hat{\mathbf{S}},
\end{equation}
where $\hat{\mathbf{S}}$ is the strain rate tensor in physical units.
Thus, we have:
\begin{equation}
\lambda = \frac{\Delta x_f}{\Delta x_c}\frac{\mathbf{S}_c}{\mathbf{S}_f}=\frac{\mathbf{S}_c}{\alpha\mathbf{S}_f}.
\end{equation}
Finally we get:
\begin{equation}
f_{i}^{neq,f} = \frac{\mathbf{S}_c}{\alpha\mathbf{S}_f}f_{i}^{neq,c}.
\end{equation}
Converting to the central-moment space by multiplying the central-moment matrix $\mathbf{M}^T$ on both sides, we have: 
\begin{equation}
\mathbf{m}_{i}^{neq,c} = \alpha\frac{\mathbf{S}_f}{\mathbf{S}_c}\mathbf{m}_{i}^{neq,f}.
\end{equation}
This is exactly the same relation expressed in Eq.~\ref{eq:m_eq_relation} for MRT model.
With the derivation in Section~\ref{sec:multi_block_formulation}, we arrive at the same mapping expression for $\mathbf{f}$ from $s_i$ to $s_j$ as given by Eq.~\ref{eq:s1_to_s2_mapping} for the non-orthogonal CMR model.

The meaning of Eq.~\ref{eq:s1_to_s2_mapping} is that when computing fluid flows with two different scales, in addition to interpolating $\mathbf{f}$ from one scale to another, we need to apply another mapping in order to obtain the correct $\mathbf{f}$, which is Reynolds number consistent.
As an illustration, take Fig.~\ref{fig:multiblock-continuous-scale} (a) for an example, which shows the setting for traditional multi-block method, where $\alpha=2$, and the mapping happens only at the coincided boundary.
The coarse scale $s_c$ iterates first before the small scale $s_f$ starts.
When $s_f$ iterates, its $\mathbf{f}$ values at the boundary, e.g., point $\mathbf{p}_f$, should first be interpolated from the nearby points of $s_c$ along the boundary, and then apply Eq.~\ref{eq:s1_to_s2_mapping} to map from $s_c$ to $s_f$, which provide the necessary boundary values for iterations at $s_f$.
After iterations at $s_f$, the $\mathbf{f}$ values at the boundary points of $s_c$, e.g., point $\mathbf{p}_c$, are first copied from the overlapped point at $s_f$, and then apply Eq.~\ref{eq:s1_to_s2_mapping} to map from $s_f$ to $s_c$ to update the boundary values of $s_c$.
We call the mapping from $s_c$ to $s_f$ as ``\textit{prior-mapping}'' and the mapping from $s_f$ to $s_c$ as ``\textit{post-mapping}''.

\subsubsection{Spatial scale mapping}
\begin{figure}[t]
	\centering
	\includegraphics[width=0.95\columnwidth]{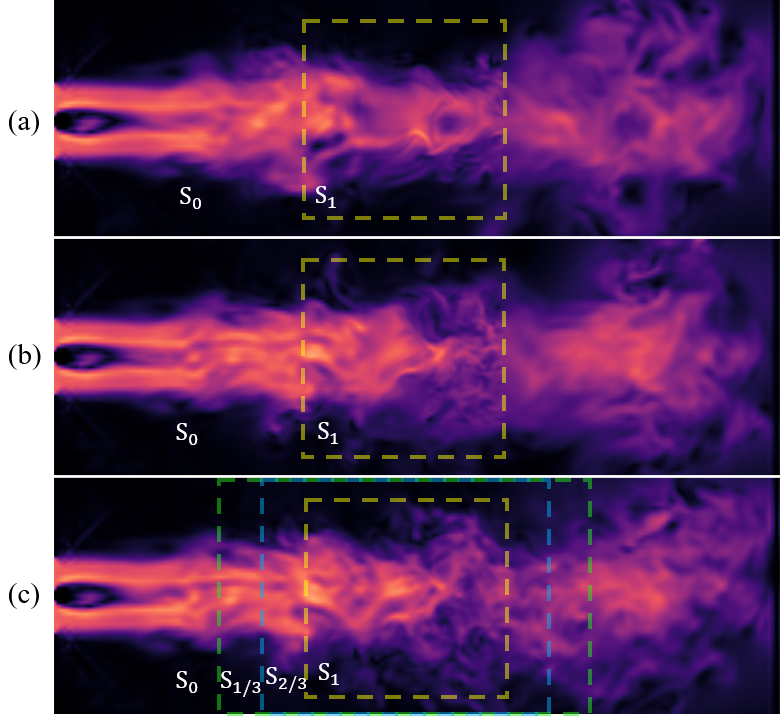}
	\caption{Comparison of transition between scales: (a) multi-block formulation (two scales) with a small portion of the overlapped region, where structure discontinuity is obvious across scales; (b) multi-block formulation (two scales) with mapping in the entire small-scale region, which reduces the blurriness but structure discontinuity still persists; (c) our continuous-scale formulation (four more continuous scales) with mapping in the entire overlapped region, which produces smooth scale structure transition with sufficient details in the small-scale regions.}
	\label{fig:multi-block-artifacts}
\end{figure}

As argued before, multi-block formulation with $\alpha$ restricted to integers is problematic especially for turbulent flows.
Fig.~\ref{fig:multi-block-artifacts} (a) shows an example of the simulation with multi-block method ($\alpha=2$) where turbulence structures are suddenly lost when transiting from fine to coarse scales due to the violation of scale-continuity.
As shown later, this discontinuity can be avoided or much reduced by constructing a more continuous-scale setting and employing our continuous-scale formulation, see Fig.~\ref{fig:multi-block-artifacts} (c), where four scales are used and turbulence structures are more continuous across the scales with details better preserved even at the coarse scale region, see the red box.

When $\alpha$ is not restricted to an integer and the boundaries between two scales do not coincide with each other (see the dark blue and red lines in Fig.~\ref{fig:multiblock-continuous-scale} (b)), we arrive at our continuous-scale setting.
In such a case, the coarse scale $s_c$ still iterates first before the small scale $s_f$, but the prior-mapping at the boundary of $s_f$, e.g., at point $\mathbf{p}_f$, requires the interpolation from the eight corner points of the 3D cell at scale $s_c$ where $\mathbf{p}_f$ locates, and then again apply Eq.~\ref{eq:s1_to_s2_mapping} to map from $s_c$ to $s_f$.
The similar procedure applies for the post-mapping at the boundary of $s_c$, e.g., at point $\mathbf{p}_c$.

\subsubsection{Transition between scales}
The descriptions above assume that the mapping between two scales only happen at the scale boundaries, which is not suitable for turbulent flows, as structure discontinuity may occur at the boundary and vortices may sometimes be blocked from going through the scale boundaries. 
To avoid these artifacts, we can extend the scales with sufficient overlaps as suggested in~\cite{Lagrava-2012}, which results in \textit{overlapped inner samples} (samples in the overlapped regions except at the boundaries), e.g., point $\mathbf{p}'_f$ in Fig.~\ref{fig:multiblock-continuous-scale} (b), where $\mathbf{f}$ values should be mapped from $s_f$ to $s_c$ after iterations at $s_f$.
\textit{In practice, we always overlay small-scales onto the large scales and overlap the entire small-scale regions}.

\begin{figure}[t]
	\centering
	\includegraphics[width=\columnwidth]{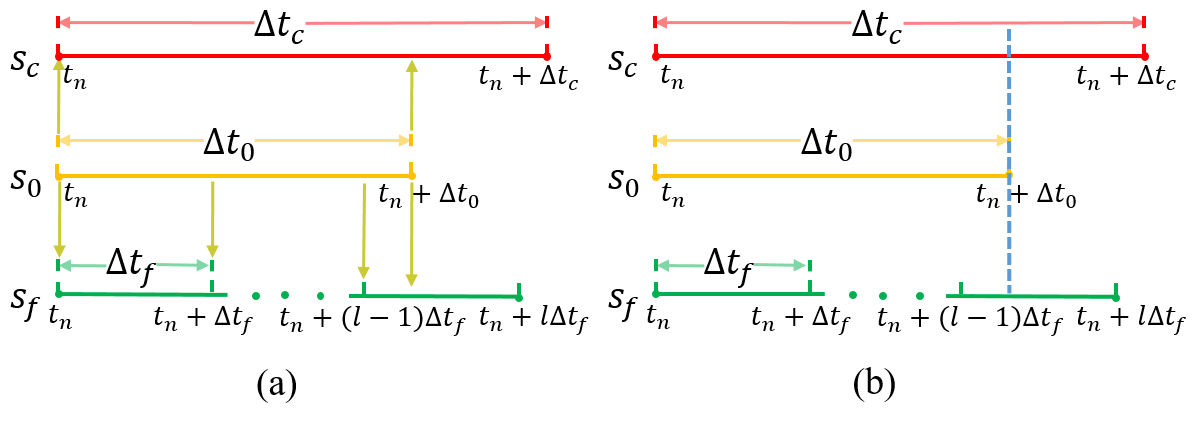}
	\caption{Temporal alignment and scale mapping in our continuous-scale formulation: (a) temporal alignment and scale mapping for all sample points at the scale boundary; (b) temporal alignment and scale mapping for overlapped inner sample points.}
	\label{fig:time-step-alignment}
\end{figure}

To justify such a treatment, Fig.~\ref{fig:multi-block-artifacts} makes a comparison, where Fig.~\ref{fig:multi-block-artifacts} (a) shows multi-block formulation with a small portion of the overlapped region along the scale boundary for mapping, which is obvious that the scale boundary blocks some flow structures from successfully going through, leading to structure discontinuity artifacts.
This can also lead to blurriness inside the small-scale region.
With mapping in the entire overlapped region, such artifacts are reduced, but cannot be removed, see Fig.~\ref{fig:multi-block-artifacts} (b).
This indicates that scale continuity can be important to preserve consistent turbulence structures across the scale boundaries.
With our continuous scale setting, we still map distribution functions in the entire overlapped region, but with more continuous scale transition (four scales rather than two), see Fig.~\ref{fig:multi-block-artifacts} (c).
It is clear that our continuous-scale setting and the related treatment can result in more consistent flow structures across different scale boundaries, making the entire fluid flow more reasonable.
Note that all the simulations in such a comparison are produced with the non-orthogonal CMR model and with our adaptive relaxation.

\subsubsection{Temporal alignment and scale mapping}
\label{sec:temporal_alignment}
Mapping distribution functions spatially only addresses the spatial consistency of local Reynolds number.
However, there are still temporal alignment and consistency problems during iterations.
To solve these problems, we should first select a reference scale $s_0$ with the reference time-step $\Delta t_0=\Delta x_0$, see the yellow time-line in Fig.~\ref{fig:time-step-alignment} (a).
Denote $\alpha_c\in(0,1)$ and $\alpha_f>1$ to be the ratios between scale $s_0$ and any other scales larger ($s_c$ with $\Delta t_c$) and smaller ($s_f$ with $\Delta t_f$) than $s_0$, respectively, see the red and green time-lines in Fig.~\ref{fig:time-step-alignment} (a).
In multi-block formulation, $s_0$ is selected as the largest scale, so we only have scale $s_f$ with $\alpha_f$ restricted to an integer.
This makes temporal alignment simple since after $\alpha_f$ iterations, scale $s_f$ naturally aligns with $\Delta t_0$.
However, in our continuous-scale setting, we can select any scale to be $s_0$, and have both $\alpha_c$ and $\alpha_f$ which are not restricted to integers.
This raises the problem that after integer number of iterations, the temporal evolutions of $s_c$ and $s_f$ may not align with $\Delta t_0$, and thus temporal interpolation is needed.
Moreover, after each iteration at scale $s_c$ and $s_f$, we need to update the boundary values from $s_0$ to maintain local Reynolds number consistency.

\begin{figure}
	\centering
	\includegraphics[width=0.95\columnwidth]{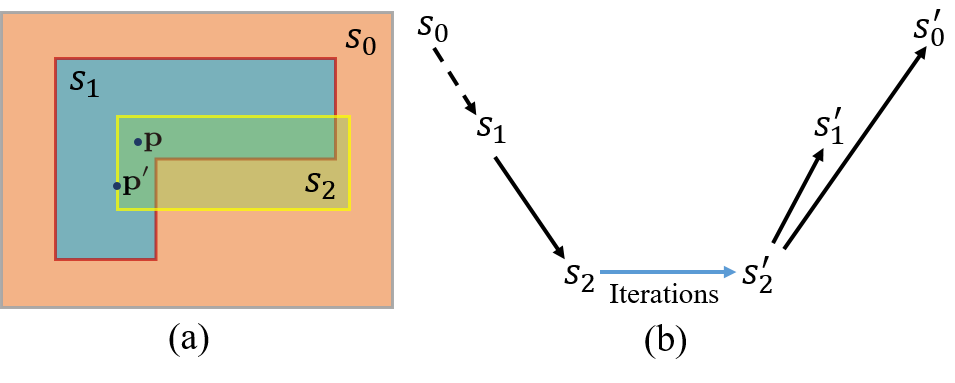}
	\caption{Handling multiple overlapped scales: (a) three scales are overlapped, where $\mathbf{p}$ is the overlapped inner sample and $\mathbf{p}'$ is the sample at the scale boundary; (b) handling scale mapping at scale $s_2$ where $s_k' (k\in\{0,1,2\})$ indicate the states of the corresponding scales after iterations. Note that the prio-mapping to $s_2$ only happens at the scale boundary while the post-mapping from $s_2$ to other larger scales happens at the entire overlapped regions.}
	\label{fig:multi-scale-handling}
\end{figure} 

To do this, note that $\Delta t_c$ and $\Delta t_f$ are determined by $\Delta t_c=\Delta x_c$ and $\Delta t_f=\Delta x_f$, respectively, which can be arbitrary.
Before any iteration of scales $s_c$ and $s_f$, we need to perform prior-mapping only on the scale boundaries except the overlapped inner samples, see the arrows at time $t_n$ in Fig.~\ref{fig:time-step-alignment} (a), which is beneficial for preserving fluid details.
To handle the overlapped inner samples for scale $s_c$ at time $t_n+\Delta t_0$, we always iterate scale $s_c$ for one time step $\Delta t_c$ and linearly interpolate back at time $t_n+\Delta t_0$ using its $\mathbf{f}$ values at $t_n$ and $t_n+\Delta t_c$, see the dotted line at scale $s_c$ in Fig.~\ref{fig:time-step-alignment} (b).
At the boundary of scale $s_c$, no temporal interpolation is needed, and their $\mathbf{f}$ values are directly mapped from scale $s_0$ at time $t_n+\Delta t_0$, see the arrow at time $t_n+\Delta t_0$ from scale $s_0$ to scale $s_c$ in Fig.~\ref{fig:time-step-alignment} (a).
For scale $s_f$ at time $t_n$, we iterate $l=\lfloor\Delta t_0/\Delta t_f\rfloor+1$ times, which may exceed $\Delta t_0$ by a fractional time-step of $\Delta t_f$.
To obtain $\mathbf{f}$ values of the overlapped inner samples at time $t_n+\Delta t_0$, we use quadratic interpolation based on the points $t_n$, $t_n+(l-1)\Delta t_f$ and $t_n+l\Delta t_f$, see the dotted line at scale $s_f$ in Fig.~\ref{fig:time-step-alignment} (b).
At the boundary of scale $s_f$, temporal interpolation of $\mathbf{f}$ values at scale $s_f$ is not needed either, and their values are directly mapped from scale $s_0$ at time $t_n+\Delta t_0$.
For the first $l-1$ iterations of scale $s_f$, $\mathbf{f}$ values at the boundary should be updated before the iteration by first interpolating from scale $s_0$ temporally using two points at $t_n$ and $t_n+\Delta t_0$, and then mapping from the interpolated values.
Before the final fractional time-step iteration of scale $s_f$ to reach the time $t_n+\Delta t_0$, we directly map $\mathbf{f}$ values from scale $s_0$ at $t_n+\Delta t_0$ to the boundary of scale $s_f$, see the arrows from scale $s_0$ to scale $s_f$ in Fig.~\ref{fig:time-step-alignment} (a).

\subsubsection{Handling multiple scales}
\label{sec:handle_multiple_scales}
In practice, there can be multiple rather than two regions with continuous scales overlapping with each other at the same point, see Fig.~\ref{fig:multi-scale-handling} (a).
To allow flexibility and adaptivity for complex domains, we construct scales such that small-scales always superimpose over large-scales with mapping in the entire overlapped regions, and the reference scale $s_0$ occupies the entire domain, which is very efficient to determine the overlapped regions between scales.
In such a case, a specific mapping scheme should be developed.

Taking point $\mathbf{p}$ and $\mathbf{p}'$ in Fig.~\ref{fig:multi-scale-handling} (a) for an example, where three scales ($s_0>s_1>s_2$) are overlapped, and $\mathbf{p}$ is the overlapped inner sample and $\mathbf{p}'$ is the scale boundary point.
Now we consider the iterations at scale $s_i$ ($s_i<s_0$).
Note that the reference scale $s_0$ should always be iterated first before any other scales $s_i$.
Prior-mapping at $s_i$ only happens at the scale boundary, so it is performed only at $\mathbf{p}'$, where $\mathbf{f}$ values should always be mapped from the nearest coarse scale $s_j, j=\text{argmin}_j \{|s_j-s_i|,s_j>s_i,j\in\Lambda(\mathbf{p}')\}$, and $\Lambda(\mathbf{p}')$ is the set of all overlapped scales at $\mathbf{p}'$.
For post-mapping, both $\mathbf{p}$ and $\mathbf{p}'$ are needed, and all the scales larger than $s_i$ ($s_k, k\in{0,1,...i-1}$) should be updated by mapping from $s_i$ directly to these scales in all the overlapped samples, but excluding the scale boundary (otherwise the fluid field may be over-smoothed), see Fig.~\ref{fig:multi-scale-handling} (b) for an example where $s_i=s_2$.
For $s_i>s_0$, the process is reversed for prior- and post-mappings.
For prior-mapping, we map all the overlapped regions from $s_0$ to $s_i$ instead; however, for post-mapping, we only update the boundaries of scales smaller than $s_i$.
The above procedure ensures reasonable continuous-scale simulations with any number of continuous scales in arbitrary overlaps, without noticeable artifacts and excessive smoothing.

Note that Zhao et al.~\cite{Zhao1-2007} employed the integral-scale multi-block approach to simulating fluid flows in graphics, with spatial interpolation along grid lines only and without any temporal interpolation.
They also applied scale mapping but with prior-mapping only.
Although our spatial and temporal interpolations may introduce a certain amount of numerical viscosity, the influence is not obvious.
In addition, since we involve post-mapping, the fluid flow information can be transferred from small scales back to large ones, and thus the simulated flows can be more turbulent.

\section{Scale construction}
\label{sec:scale_construct_tracking}

\begin{figure}[t]
	\centering
	\includegraphics[width=0.95\columnwidth]{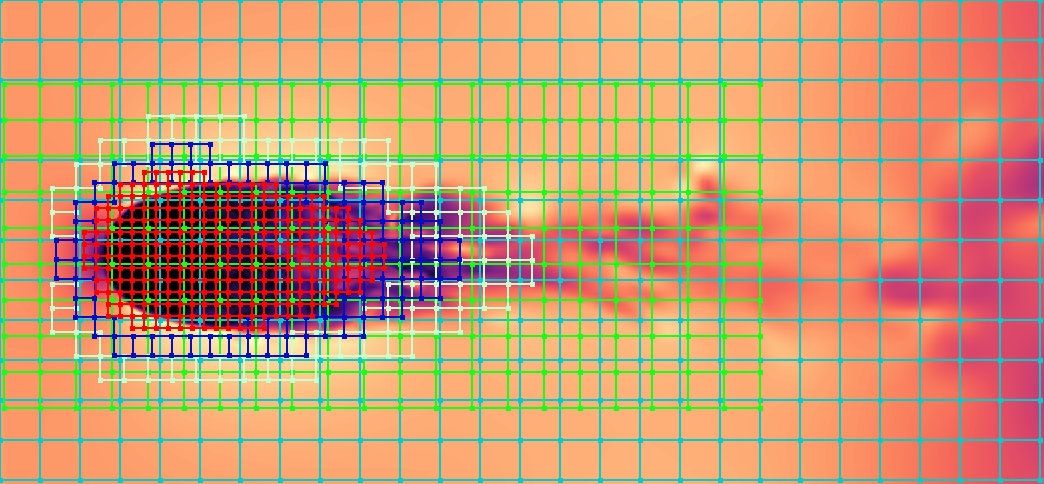}
	\caption{Static scale construction: The scale samples at initialization are constructed according to the domain geometry as well as the inlet position. In this example, since the solid ball is the object boundary, very dense samples are placed there. In addition, more samples are placed in the wake flow region in order to capture the wake turbulence.}
	\label{fig:static_scale}
\end{figure}

\begin{figure*}[t]
	\centering
	\includegraphics[width=0.95\textwidth]{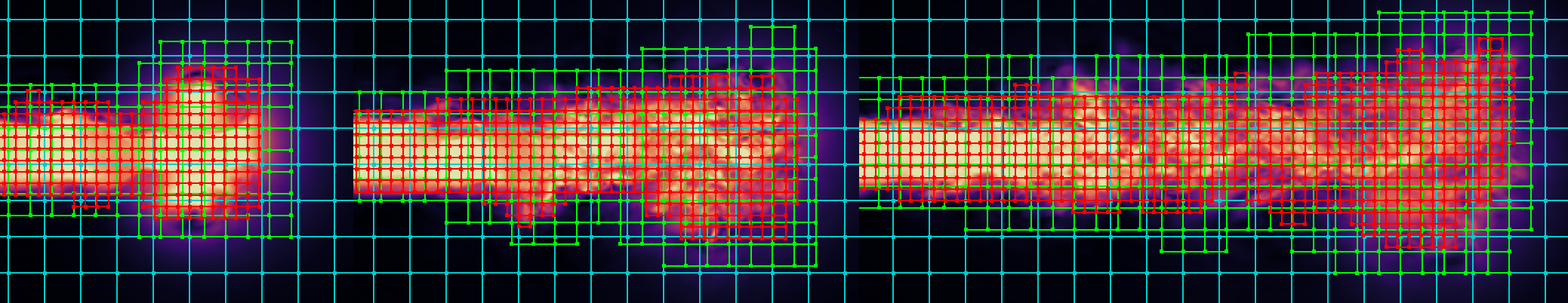}
	\caption{Dynamic scale construction for a jet flow: Based on the initial static scale samples, the fine scale samples with higher resolutions can be dynamically overlaid onto all the other scales, which capture fine turbulence details only when necessary. In this example, two dynamic scales are created, which track the evolution of the turbulence, see the green and red samples.}
	\label{fig:dynamic_scale}
\end{figure*}

The continuous-scale formulation proposed in Section~\ref{sec:contineous-scale-formulation} immediately allows us to have flexible scale construction since it looses the geometrical restriction between scales.
In principle, two types of scale construction schemes can be used for simulation.
During initialization, \textit{static scale construction}, where the scales are purely determined by domain geometries (the shapes of objects inside the fluid region and the domain boundary) and inlet positions, is performed to place finer scale samples around object boundaries and the wake-flow regions behind them, as well as coarser scale samples for fields far from the simulation domain.
During the simulation, \textit{dynamic scale construction}, which refines the scales over time, adapts scale samples dynamically to track the turbulence details.

\begin{algorithm}[t]
	\caption{Pseudo-code for our continuous-scale kinetic fluid simulation method}
	\begin{algorithmic}
		\STATE Initialize density $\rho$ and velocity $\mathbf{u}$ for all scales;
		\STATE Initialize distribution functions $f_i$ with their equilibrium states according to Section~\ref{sec:initial_boundary_treatment};
		\STATE Define the reference scale in DOI and scales in FFD according to Section~\ref{sec:scale_construct_tracking};
		\newline
		\WHILE {iteration $\leq$ max iteration number}
		\STATE Proceed one time-step for the reference scale and scales in FFD with boundary conditions specified in Section~\ref{sec:initial_boundary_treatment};
		\STATE Construct dynamic scales and overlay them on all other scales (optional);
		\STATE Set current scale = reference scale;
		\WHILE{current scale $\geq$ smallest scale}
		\STATE Proceed to the next overlapped smaller scale;
		\STATE Perform prior-mapping on current scale from the overlapped nearest larger scale according to Section~\ref{sec:handle_multiple_scales};
		\STATE  Proceed several time-steps for the current scale to match the temporal boundary of the reference scale according to Section~\ref{sec:temporal_alignment} with boundary updates;
		\STATE Update the overlapped scales larger than the current scale according to Section~\ref{sec:handle_multiple_scales};
		\ENDWHILE
		\ENDWHILE
	\end{algorithmic}
	\label{tab:lbmalgo}
\end{algorithm}

\vspace{0.1cm}
\noindent
\textbf{Static scale construction.}
To construct static scales, we rely on the distance map of the domain geometry, where each point in the domain is given the shortest distance to the object or domain boundaries, which is an indicator of scales, where smaller distance implies smaller scales.
Such a distance field is then quantized and mapped to $N$ discrete scales, where $N$ is manually determined (e.g., $N=5$).
To decide the exact spacing for each scale, we assign the smallest and largest spacings for the smallest and largest scales respectively, and linearly interpolate the spacings for scales in between.

In case of wake flows where turbulence could be strong, we determine the scales like a soft shadow in rendering, where the inlet is taken as an area light source, and the scale becomes small when a point is in the ``umbra'' region behind the object and gradually increases for points in the ``penumbra'' region.
In addition, for practical simulations where the open space is usually configured, we divide the whole simulation domain into domain of interest (DOI) and far-field domain (FFD), like in~\cite{Zhu-2013}.
In DOI, the largest scale is taken as the reference scale, and in FFD, scales are increased from the boundary of DOI and becomes very large at its outer boundary to damp out turbulent variations to approximate a real open space.

Fig.~\ref{fig:static_scale} demonstrates an example of static scale construction for flows around a solid ball, where five static scales are created.
As mentioned in Section~\ref{sec:contineous-scale-formulation}, we always overlay the entire regions of small-scales over large-scales.
With such a setting, boundary induced turbulence can be better resolved.
Note that the samples in Fig.~\ref{fig:static_scale} are for illustration only; the true number of samples is several times denser.

\vspace{0.1cm}
\noindent
\textbf{Dynamic scale construction.}
The dynamic scales are fine scales that are dynamically created and overlaid onto all the other scales, which may change over time, and are beneficial to have necessary computations only when needed.
The dynamic scales are more complicated to construct and sophisticated methods are required to create sufficiently connected and large enough regions.
In this paper, we develop a simple dynamic scale construction method particularly for jet flows, which are very efficient to compute.

To create such scales, we first compute the gradient magnitude of the entire velocity field.
Then we select a threshold to remove samples with the gradient magnitude below such a threshold.
This is to ensure that the dynamic scales are constructed only in sufficiently fluctuating regions to capture the turbulent flow details.
For the remaining samples, we threshold again, but instead based on the velocity magnitude, to further remove samples that have velocity magnitude smaller than the threshold.
This is to ensure that the dynamic scales are constructed with sufficient spatial continuity.
After these two thresholding processes, the remaining samples form one dynamic scale region.
In practice, multiple thresholds can be selected to create multiple dynamic scales, and such a process is repeated for every $40$ iterations in our practical simulations.

Fig.~\ref{fig:dynamic_scale} shows a typical dynamic scale construction process for a jet flow, where two time-varying dynamic scales are constructed, see the green and red samples, where sufficient flow details can be captured inside these regions.
As can be seen in the timing statistics later in the next section, the dynamic scale construction effectively saves computations to solve for turbulent flows, which occupy only a portion of the entire flow region.

\section{Results and Discussions}
\label{sec:results_discussions}

\begin{figure*}[t]
	\centering
	\includegraphics[width=\textwidth]{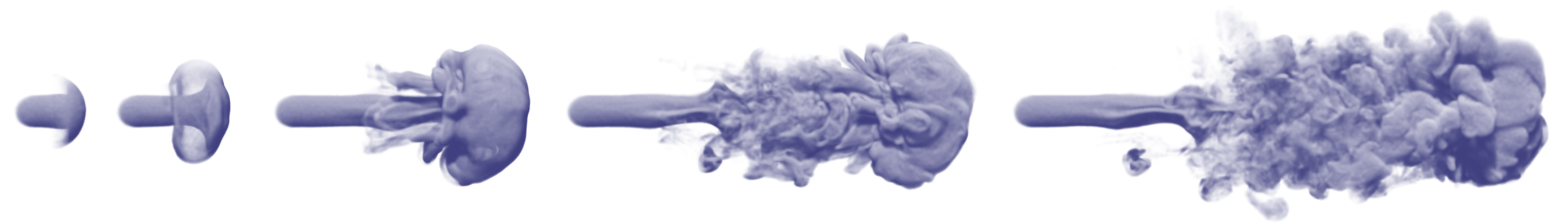}
	\caption{Smoke simulation without obstacles: the air flow with smoke particles is injected from the left side of the domain and evolves to the right. In order to capture fine turbulence details while reducing the computational cost, we employ dynamic scales which are overlaid onto the static scale and evolve over time. The smoke is rendered with tracing particles.
	}
	\label{fig:smoke_result1}
\end{figure*}

\begin{figure*}[t]
	\centering
	\includegraphics[width=\textwidth]{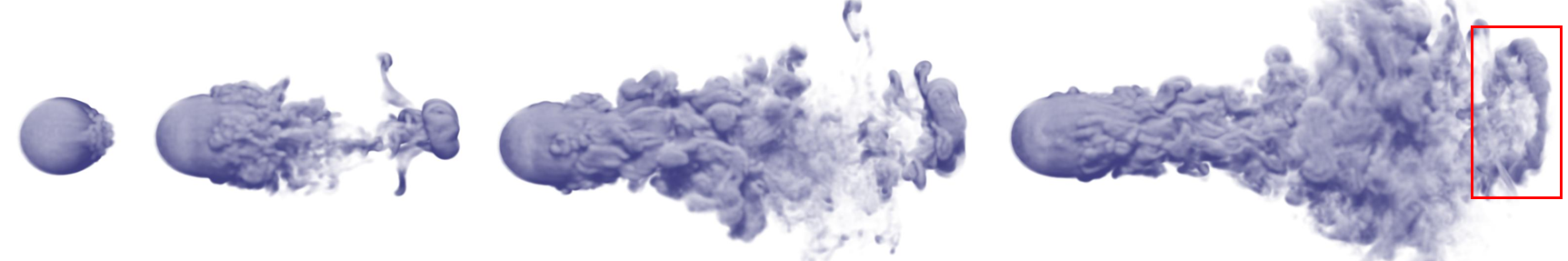}
	\caption{Smoke simulation with a ball obstacle: the air flow is injected from the left side of the domain and evolves to the right. The smoke source is placed around the surface of the ball. In order to capture fine turbulence details, more samples with continuous scales are placed around the ball surface as well as the wake-flow region behind the ball, which captures the fine details of the smoke. In addition, a vortex ring, as indicated by the red box, is produced and evolves over a long time.
	}
	\label{fig:smoke_result2}
\end{figure*}

With our continuous-scale formulation and scale construction methods, we can realize our fluid simulator, and Algorithm~\ref{tab:lbmalgo} gives the pseudo-code as a reference for implementation.
Note that when implementing our method, we need to first select a domain with a physical size which could be arbitrary, and then determine the physical spacing $\Delta \tilde{x}_i$ for each scale, which is finally used to determine the overlap and resolution of each scale.
Note that in our kinetic simulator, $\Delta x$ is normalized and always set to 1; when we specify different scales, we use $\Delta \tilde{x}_i$ instead.  

We implement our simulator on a computer installed with an Intel Xeon E5-2630 v3 @2.4GHz CPU and 48 GB system memory.
Our method is easily parallelizable and the main iterations are implemented on an NVIDIA GTX 1080 GPU with 8 GB onboard memory, where our simulations take from $2$ to $7$ GB memory with overall number of samples from around $1.5\times10^6$ to $6.8\times10^6$ depending on the scenario we simulate.
For each iteration of the entire domain to finish with respect to the reference scale including the FFD scales, our method takes around $0.4$ to $2.3$ seconds, without rendering.
To visualize the velocity field, we take a cross-section and use direct color-mapping.
Note that to produce one animation frame, we usually have around 10 iterations.

\subsection{Initialization and boundary treatment}
\label{sec:initial_boundary_treatment}

To initialize the fluid flow field, we first give a constant initial density $\rho_0 = 1$, and a calm velocity field $\mathbf{u}_0 = 0$ except at the inlet where the velocity is set as $\mathbf{u}_0 \in [0.1,0.13]$, which is also taken as the Dirichlet boundary to keep injecting the flow into the domain. 
These macroscopic fields are then converted to distribution functions for each sample based on the equilibrium state of the CMR model: $\mathbf{f}_0=\mathbf{T}\mathbf{m}^{eq}(\rho,\mathbf{u})$.
For boundaries around objects including the ground, we apply a second-order no-slip boundary treatment method described in \cite{Latt-2007}, which is more accurate.
Note that the traditional no-slip bounce-back boundary treatment is first-order accurate only and cannot give stable simulations.
For FFD boundary, we use the Neumann condition.
It should be noted that although standard boundary conditions (slipping and no-slip conditions) can be easily applied in our solver, some particular boundary conditions should be further derived and may not be straightforward to apply, which can be a potential drawback of the kinetic approach for simulating fluid flows in more complex environments.

\subsection{Stability and accuracy}
With our adaptive relaxation in non-orthogonal CMR model, stability and accuracy can be retained, which allows very small viscosities (e.g., $10^{-6}$) or even zero viscosity with small vortices, and with arbitrary boundary geometries.
Our method does not rely on any turbulence model for stabilization, which reduces the uncertainty during simulations.
As an example, Fig.~\ref{fig:vortex_ring} shows a rotating vortex ring from the boundary layer induced smoke flow with rotating structures around the ring, which can be generated only with accurate advection solvers especially at the boundary, and could be preserved with our method for a long time even in regions with coarse resolutions. 
However, the transition between scales rely on interpolation, which may break the conservation of the original LBE and introduce a certain amount of numerical diffusion that may smear out small-scale details.

When we say our solver is more accurate than the traditional ones, we mean it can faithfully preserve the necessary small-scale structures for more visual realism without the aid of any other empirical models.
From the computational side, the model accuracy is reflected in two aspects: i. the collision model responsible for approximating INSE is more accurate, where non-orthogonal CMR model with our adaptive relaxation greatly reduces the ghost modes and ringing artifacts, and has a higher approximation order to the corresponding INSE; ii. the discretization on both space and time are second order including the boundary treatment, with conservative advection, which are also important to preserve turbulence structures.
The main influence on discretization accuracy is the interpolation when mapping among different scales, but not obvious in practice.

\subsection{Parallel implementation}
Since our method is local in dynamics, it is easy to be parallelized on the GPU for fast computation.
Since scales are coupled in overlapped regions in our method, we start from the reference scale and progress gradually to the smaller and larger scales in a serialized order.
However, at each scale, we iterate the dynamics in parallel.
If one scale has multiple regions, these regions are also iterated in parallel.
Note that in our parallel implementation, we do not have any code or hardware-level optimizations, and the timings presented later are based on such a straightforward implementation, which could be further improved in the future.

\subsection{Smoke simulation}
To demonstrate the applicability, we apply our method to smoke simulations, where we inject around $2,000$ to $5,000$ smoke particles per iteration into the domain from a user specified smoke inlet (note that the smoke inlet can be different from the velocity inlet, where we ignore combustion).
The particles move by integrating their positions with respect to the velocity field by 3rd-order Runge-Kutta method~\cite{Lomax-1999}, and are rendered with particle renderer from \cite{Zhang-2015}, where the average rendering time is 6 to 40 seconds on the CPU per animation frame (we iterate around 10 times to generate one such frame) depending on different scenarios.
Our scale construction can easily allow dense samples to be placed around the complex object with sufficient flexibility, which improves the accuracy of the important flow field, and thus more plausible visual effects around complex object boundaries can be obtained.

\begin{figure}
	\centering
	\includegraphics[width=\columnwidth]{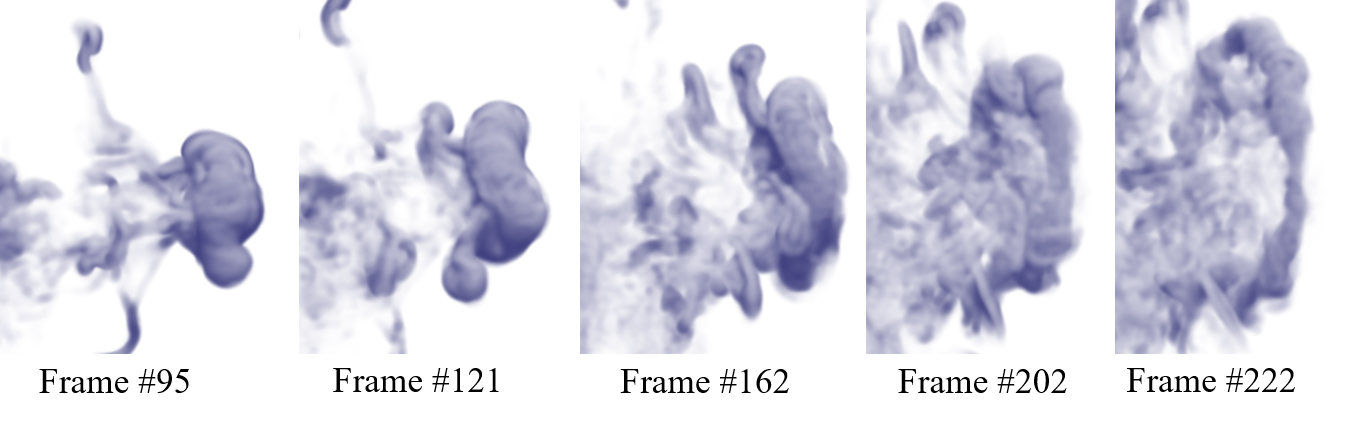}
	\caption{Vortex ring in our simulated boundary layer induced smoke. The smoke vortex ring is generated at the very early stage of the simulation due to the shearing of the boundary layer. But it evolves and is well preserved by our solver for a long time, even at the region which is four times sparser in sample resolution than the region around the ball.}
	\label{fig:vortex_ring}
\end{figure}

Fig.~\ref{fig:smoke_result1} shows the snapshots of a smoke simulation with an inlet in the left of the domain.
In order to capture sufficient details and reduce storage and computation cost, we employ dynamic scales which are overlaid onto the reference scale and are evolving over time.
The spacing for the reference scale is $\Delta \tilde{x}_0=3$ and the spacings for the dynamic scales are $\Delta \tilde{x}_i=\{1.8, 1.1\}$.
No FFD scales are used.
There are $1.5\times10^6$ samples used in the simulation with $4$ seconds on average to produce one animation frame, and the maximum memory cost is 3 GB on both CPU and GPU.
It is clear that the fine details can be well preserved.

While Fig.~\ref{fig:smoke_result1} shows a smoke simulation without any object inside, Fig.~\ref{fig:smoke_result2} shows the smoke simulation where a ball object is placed inside the domain and the smoke is injected from the surface of the ball.
In such a case, we use static scales only and place dense samples around the ball as well as its wake-flow region.
The reference scale is $\Delta \tilde{x}_0=2$; the spacing for the FFD scale is $\Delta \tilde{x}_i=4$; and the rest static scales are: $\Delta \tilde{x}_j=\{1.4, 1, 0.8, 0.5\}$.
There are $4.8\times10^6$ samples used in the simulation with $12$ seconds to produce one animation frame, and the memory cost is 5 GB on both CPU and GPU.
It is clear that boundary layer turbulence details can be well preserved.
It can also be noticed that a clear rotating vortex ring is produced and maintained for a long time, see Fig.~\ref{fig:vortex_ring}.
Note that only high resolution simulation with sufficient accuracy around the ball can produce such a vortex ring.
Due to our solver flexibility to place higher resolution samples near the ball and thus higher accuracy, the vortex ring can be well generated, and due to conservative advection with more accurate collision model in each scale in our method, the vortex ring can be preserved without over smoothing even in the region with coarse grid resolution (the right region of the domain) which is four times sparser than the resolution around the ball.

\begin{figure*}[t]
	\centering
	\includegraphics[width=\textwidth]{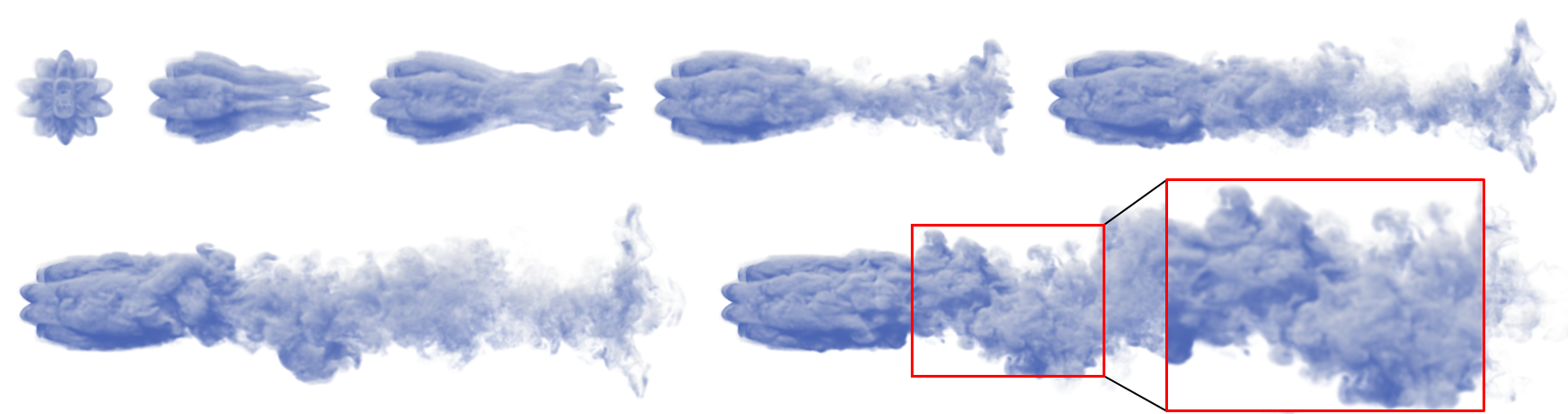}
	\caption{Boundary layer turbulence of a complex object. The smoke is injected near the surface of the object, and then follows the wake flow to generate complex turbulence patterns. Note the smoke details as illustrated in the red box.}
	\label{fig:smoke_result3}
\end{figure*}

\begin{figure*}[t]
	\centering
	\includegraphics[width=\textwidth]{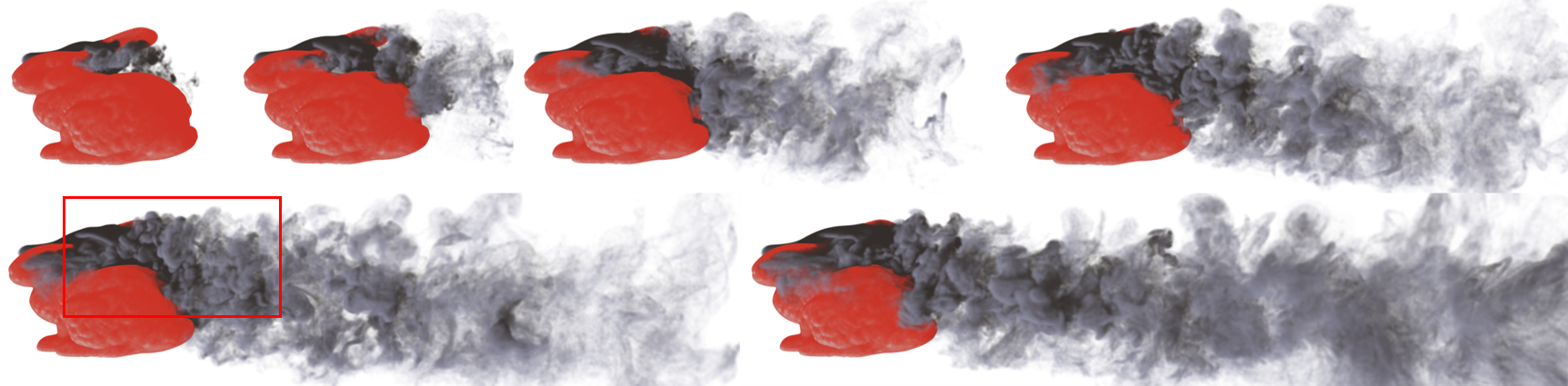}
	\caption{Smoke simulation over a bunny object which is of complex shape. The smoke starts from a source behind the front ear of the bunny model. Note that due to the accuracy of our solver to address boundary layer flows more appropriately, the vortices around the concave region produce complex swirling motion, see the red rectangle box region and the related animation in the supplementary video.}
	\label{fig:smoke_result4}
\end{figure*}

The more powerful capability of our solver is to tackle arbitrary geometrical boundaries in an efficient manner.
Figs.~\ref{fig:smoke_result3} \& \ref{fig:smoke_result4} give two examples of air flows passing through objects with complex shapes, where smokes are injected near the boundary.
For clarity, we only inject smokes over the small area near the object in Fig.~\ref{fig:smoke_result4} .
In Fig.~\ref{fig:smoke_result3}, the spacing for the reference scale is $\Delta \tilde{x}_0=2$ and we use two FFD scales which are $\Delta \tilde{x}_i=\{3, 4.5\}$; the spacings for the rest static scales are: $\Delta \tilde{x}_j=\{1.1, 1, 0.8, 0.5\}$, with CPU and GPU cost of 6.2 GB for $5.7\times10^6$ samples, and it takes $16$ seconds to produce one animation frame.
In Fig.~\ref{fig:smoke_result4}, the spacing for the reference scale is $\Delta \tilde{x}_0=2$ and we use two FFD scales which are $\Delta \tilde{x}_i=\{3, 4.5\}$; the spacings for the rest static scales are: $\Delta \tilde{x}_j=\{1.6, 1, 0.7, 0.5\}$, with CPU and GPU cost of 5.3 GB for $5\times10^6$ samples, and it takes $13$ seconds to produce one animation frame.
It is clear that visually appealing smoke patterns can be produced.
\textit{Readers are suggested to refer to the supplementary video for these animations}, and for smoke motion in Fig.~\ref{fig:smoke_result4} and in the related animation, it can be observed that the swirling feature of the smoke due to concave geometry can be faithfully resolved.

\subsection{Comparisons}
\label{sec:comparison}

\begin{figure*}[t]
	\centering
	\includegraphics[width=1\textwidth]{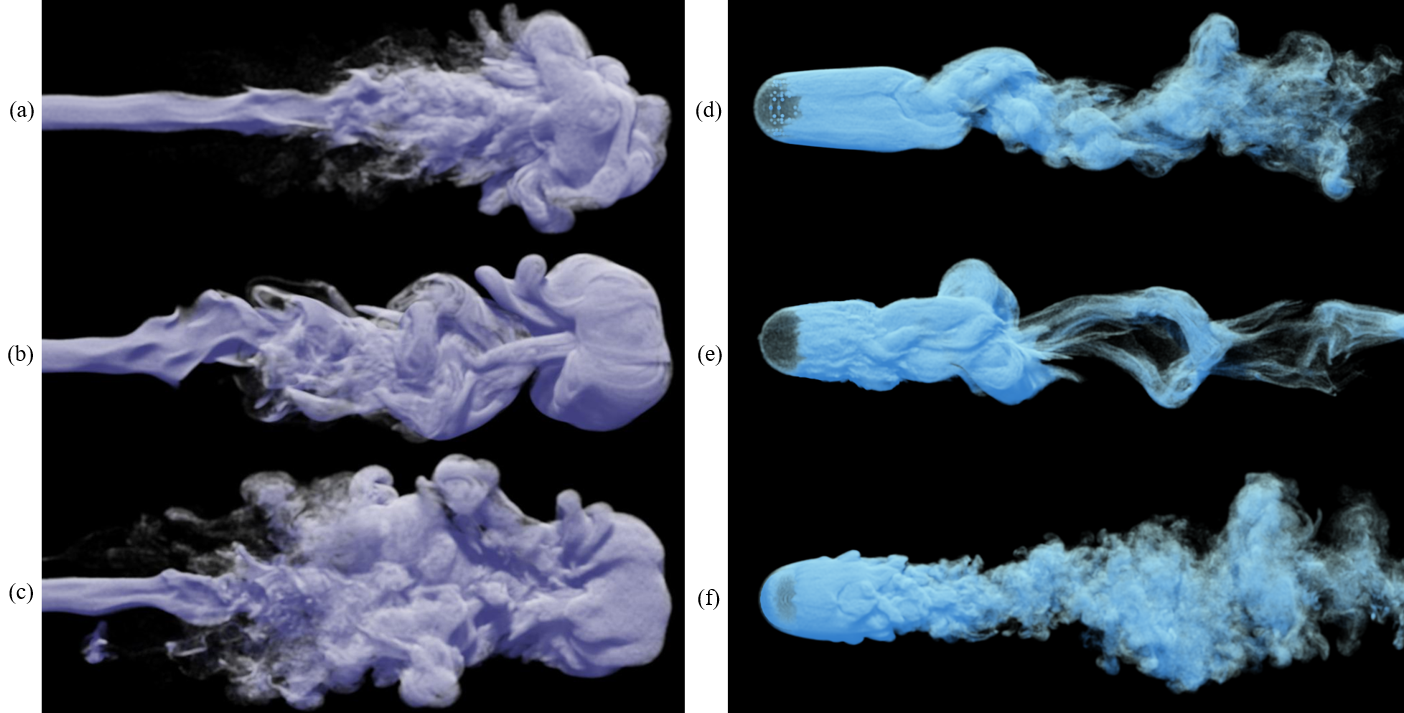}
	\caption{Comparisons of smoke simulations with the unconditionally stable MacCormack scheme (a) \& (d) and recent work of Zhang et al.~\cite{Zhang-2015,Zhang-2016} (b) \& (e).
		In (a) \& (d), the incompressible Euler equation is solved with the method of unconditionally stable MacCormack scheme for flows without and with objects, while in (b) \& (e), the flows with the same setting are solved using IVOCK scheme from \cite{Zhang-2015} and adaptive grid method from \cite{Zhang-2016}, respectively.
		We simulate the corresponding flows with our solver and with the same setting to produce results in (c) \& (f).
		In (c), we set a very small viscosity to approximate the incompressible Euler equation, while in (f), we set the same viscosity as in (e).
		It is clear that more appropriate turbulence details can be resolved with our method, especially around the object in (f).
	}
	\label{fig:comparison}
\end{figure*}

To verify our method, we conduct comprehensive comparisons with the well-known unconditionally stable MacCormack method~\cite{Selle-2008} as well as the more recent work from Zhang et al.~\cite{Zhang-2015,Zhang-2016} for smoke simulations, where similar initial and boundary conditions as well as averaged resolutions are used between their simulations and ours, see Fig.~\ref{fig:comparison}.
\textit{Note that the combustion force in the original simulations of these existing methods is ignored in all following comparative simulations in order to demonstrate and compare the capability of capturing self-initiated turbulence without external activation}.
We also compare boundary-induced turbulence under different resolutions to highlight the advantage of flexibility for our method, see Fig.~\ref{fig:single_adpative_compare}.
\textit{Readers are suggested to see the supplementary video for animations of these comparisons.}

\vspace{0.1cm}
\noindent
\textbf{Comparisons with MacCormack advection scheme.}
In Fig.~\ref{fig:comparison} (a) \& (d), we simulate jet flows and boundary layer induced smoke motion respectively both by solving the incompressible Euler equation using the well-known second-order unconditionally stable MacCormack advection scheme, which is the standard approach for smoke simulations in graphics.
In Fig.~\ref{fig:comparison} (a), $2\times10^6$ samples ($100 \times 200 \times 100 $) are used to obtain the result.
Since the advection is not accurate enough, turbulence is activated very late, and less small-scale vortices are created, which makes the simulation not quite realistic.
With the same setting, we obtain our simulation result in Fig.~\ref{fig:comparison} (c), where we use only $1.5\times10^6$ samples in the maximum case with three scales (the reference scale is $\Delta \tilde{x}_0=3$ and the other scales are $\Delta \tilde{x}_i=\{1.8,1.1\}$, without FFD scales) to simulate the smoke motion since we apply dynamic scales in this simulation.
Since our method solves INSE, we use a very small viscosity ($10^{-5}$) to approximate the result, which leads to a Reynolds number of $10^6$.

Both the simulations are implemented on the same GPU, where the unconditionally stable MacCormack scheme with preconditioned conjugate gradient (PCG) pressure solver takes around 3 seconds with 2 iterations to produce one animation frame while our method takes around 4 seconds with $10$ iterations.
Note that for unconditionally stable MacCormack scheme, we can use larger CFL number and we choose CFL=3 to balance between efficiency and accuracy.
Although our method is slower than this traditional scheme, more reasonable turbulence patterns can be generated for our method in both horizontal and vertical directions, making our simulated smoke motion more plausible.
The memory usage for Fig.~\ref{fig:comparison} (a) \& (c) are 0.4 GB on average and 3 GB in the maximum, respectively. 

We also simulate the boundary layer induced smoke motion by solving the incompressible Euler equation with with around $4\times10^6$ samples ($128 \times 256 \times 128 $) using the unconditionally stable MacCormack scheme, with boundary treatment method by~\cite{Bridson-2015}, which is shown in Fig.~\ref{fig:comparison} (d).
In Fig.~\ref{fig:comparison} (f), we obtain our simulation result with the same setting using $3.8\times10^6$ samples and $7$ number of static scales (the reference scale is $\Delta \tilde{x}_0=2$ and FFD scales are $\Delta \tilde{x}_i=\{3, 5\}$; the other scales are $\Delta \tilde{x}_j=\{1.5, 1, 0.8, 0.5\}$).
Both simulations are also implemented on the same GPU, where it takes around 9.8 seconds with $4$ iterations for Fig.~\ref{fig:comparison} (d) to produce one animation frame with CFL=5, while our method takes $10$ seconds with $15$ iterations to produce result in Fig.~\ref{fig:comparison} (f).
The memory usages for Fig.~\ref{fig:comparison} (d) \& (f) are 1 GB and 4 GB, respectively. 
Unlike the previous comparison, the unconditionally stable MacCormack scheme in this boundary layer simulation case has almost similar computational time on the GPU as our method, but unable to capture sufficient turbulence details.

\begin{figure}[t]
	\centering
	\includegraphics[width=1\columnwidth]{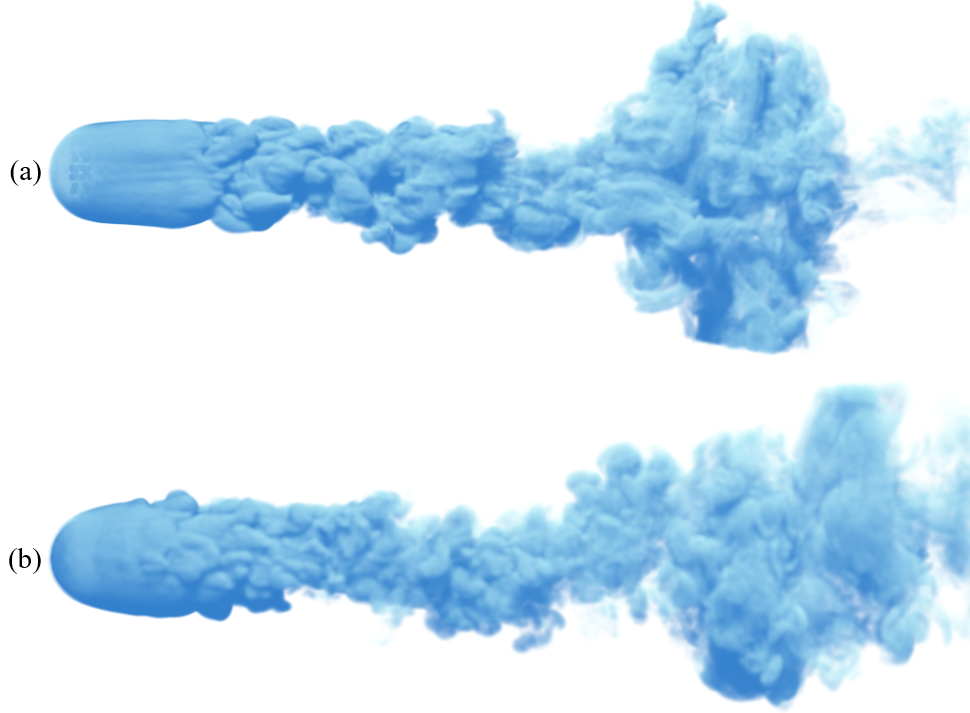}
	\caption
	{
		Comparison of smoke simulations between high resolution unconditionally stable MacCormack scheme (four times higher in resolution than Fig.~\ref{fig:comparison} (d), around $1.6\times 10^{7}$ samples) and our method. It is clear than high resolution MacCormack scheme produces more vortices than Fig.~\ref{fig:comparison} (d) and is closer to our method, but it also costs much more computing time and more memory usage than that in Fig.~\ref{fig:comparison} (d).
	}\label{fig:comparisonHighMac}
\end{figure}

To further capture fine turbulence details for the unconditionally stable MacCormack scheme, we can either reduce CFL number or increase the sample resolution.
Both methods will significantly increase the required computing time.
For example, Fig.~\ref{fig:comparisonHighMac} makes a comparison of simulation results between our method (Fig.~\ref{fig:comparisonHighMac} (b)) in Fig.~\ref{fig:comparison} (f) ($3.8\times10^6$ samples)
and higher resolution unconditionally stable MacCormack scheme (Fig.~\ref{fig:comparisonHighMac} (a), four times higher than the simulation in Fig.~\ref{fig:comparison} (d) with $1.6\times10^7$ samples).
It can be seen that Fig.~\ref{fig:comparisonHighMac} (a) produces more turbulence details than Fig.~\ref{fig:comparison} (d) and is closer to our simulation result, but it also takes much more computing time on GPU (35 seconds per animation frame) with $3.5$ GB memory usage. 
Thus, the advantage of our method is obvious.

\vspace{0.1cm}
\noindent
\textbf{Comparisons with methods of Zhang et al.}
To preserve vortices and retain more turbulence details with the same CFL number and sample resolution, the unconditionally stable MacCormack scheme can be enhanced by the IVOCK scheme proposed Zhang et al.~\cite{Zhang-2015}, which is demonstrated in Fig.~\ref{fig:comparison} (b), where $2\times10^6$ samples ($100 \times 200 \times 100 $) are used to obtain the simulation result, which is equal to the number of samples in Fig.~\ref{fig:comparison} (a).
It is clear that the turbulence structures are more reasonable compared to the result from the unconditionally stable MacCormack scheme in Fig.~\ref{fig:comparison} (a), but some small-scale details are still missing.
In addition, due to the involvement of vorticity correction for IVOCK scheme, much more computation is added to the solver.
Since the IVOCK scheme is more difficult to have a GPU implementation, we run it on a CPU with $2.3$ GHz frequency, but with parallel execution on 20 cores to maximize its performance.
It costs around $50$ seconds with 2 iterations and fixed CFL number (CFL=3) for the method of Zhang et al.~\cite{Zhang-2015} to produce one animation frame, which costs 2 GB memory, while our solver costs around $4$ seconds, which is much faster, but with more memory (4 GB).
If we offload the IVOCK scheme to the GPU and considering that a GPU implementation of an incompressible Euler equation solver is generally only several times faster than its CPU equivalent, our solver is still very promising in performance while capturing even more turbulence details than the IVOCK scheme.

To enhance the simulation for boundary layer induced smoke motion, we employ the method of Zhang et al.~\cite{Zhang-2016} for comparison, and Fig.~\ref{fig:comparison} (e) shows such a simulation with $3.7\times10^6$ samples (two scales), which is almost equivalent to the number of samples in our simulation in Fig.~\ref{fig:comparison} (f).
To increase turbulence details, we set zero viscosity for Fig.~\ref{fig:comparison} (e) and a viscosity of $10^{-4}$ for Fig.~\ref{fig:comparison} (f), which results in a Reynolds number of $5\times10^5$.
From the comparison, it is clear that our solver can capture much more boundary layer turbulence details, especially in the wake flow region.
Since the solver by Zhang et al.~\cite{Zhang-2016} is also more difficult to be implemented on the GPU, we still run it on a CPU with $2.3$ GHz frequency, but with 20 cores for parallel execution to maximize its performance, and it costs around $80$ seconds with $1$ iteration for the method of Zhang et al.~\cite{Zhang-2016} to produce one animation frame with 3 GB memory, while our solver costs around 10 seconds and 4 GB memory, with $15$ iterations per animation frame.
Such a solver can be implemented on the GPU, and as argued similarly before, the performance can only be several times faster than its CPU equivalent.
In this case, our solver is also very promising in performance for boundary layer flow simulations.

\vspace{0.1cm}
\noindent
\textbf{Comparison under different resolutions.}
In addition to the comparisons above, it is also interesting to compare our solver under different resolutions, where
Fig.~\ref{fig:single_adpative_compare} (a) \& (c) show the smoke simulations passing though a ball, which are solved on uniform grids only without any adaptive refinement, but with higher (around $6\times10^6$ samples in DOI region in Fig.~\ref{fig:single_adpative_compare} (a)) and lower (around $1\times10^6$ samples in DOI region in Fig.~\ref{fig:single_adpative_compare} (c)) resolutions respectively, while Fig.~\ref{fig:single_adpative_compare} (b) shows the simulation result with our continuous-scale setting, but with higher resolution near the ball and the total number of samples ($3.5\times10^6$ samples in DOI region) is almost in the middle between those in Fig.~\ref{fig:single_adpative_compare} (a) \& (c). 

It is clear that with our continuous-scale setting and flexible sample placement around object boundary (the resolution around the ball is even higher than the one in Fig.~\ref{fig:single_adpative_compare} (a)), more plausible turbulence structures can be captured around the ball, which is closer to the high resolution result in Fig.~\ref{fig:single_adpative_compare} (a), with even finer details.
Such a turbulence structure can also be transmitted to the wake flow region far behind the ball, where we use a very coarse grid (four times sparser than the region around the ball), but the fluctuation can also be well preserved without significant diffusion due to conservative advection and collision in each scale.
Considering that almost half number of samples are used in our simulation to produce a result even better than the one with uniform high resolution grid, our method obviously has performance gains. 

\begin{figure}[t]
	\centering
	\includegraphics[width=\columnwidth]{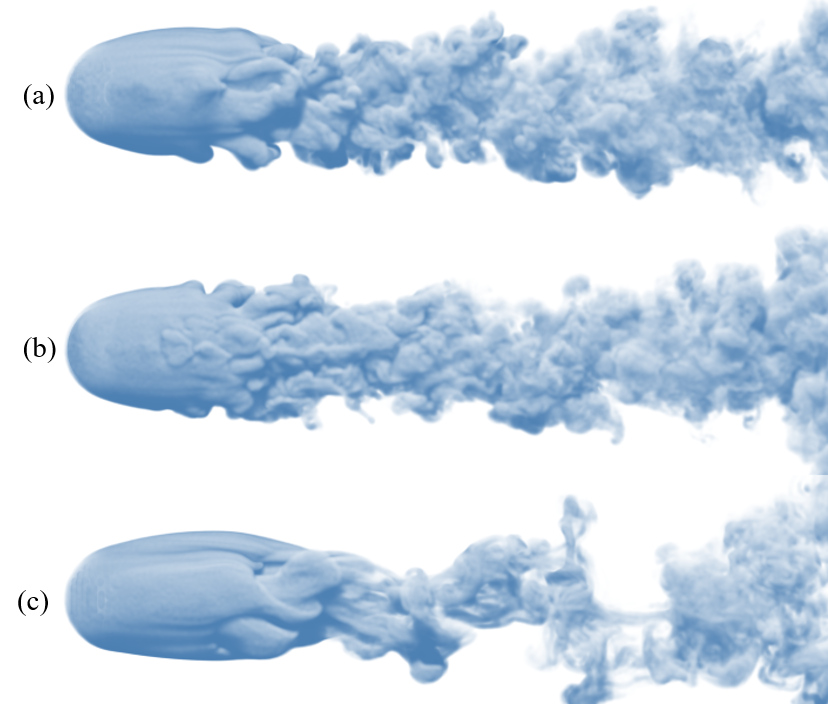}
	\caption{Comparison of flows passing through a ball with different resolutions and sample placement. (a) uniform grid with high resolution (around $6\times10^6$ samples in DOI region); (b) our adaptive resolution with much less number of samples ($3.5\times10^6$ samples in DOI region) and thus less computation time, but higher resolution around the ball; (c) uniform grid with low resolution (around $1\times10^6$ samples in DOI region). It is clear that, our simulation produces closer vortex structures around the ball to the high resolution simulation, with even finer details, which is much better than the low resolution simulation result.}
	\label{fig:single_adpative_compare}
\end{figure}

\subsection{Advection efficiency}
In traditional kinetic approach with lattice Boltzmann method, the advection is under the restriction of CFL=1, which does not allow flexible tuning of time steps, and may first seem to be slow. 
However, considering the balance between efficiency and accuracy, we usually do not set a very large CFL number for traditional macroscopic solvers, and the maximum CFL number is usually set around 3 or even smaller. 
In this case, kinetic solver is not really slow, especially considering the adaptive continuous scale setting where the reference scale is usually much coarser than the smallest scale, resulting in relatively large time steps for one iteration of the entire domain.
The parallel implementation is also straightforward, and the whole solution is almost conservative, which is beneficial for turbulent flows.
The comparisons in Section~\ref{sec:comparison} verified above arguments.

\subsection{Limitations}
Our method also suffers from several limitations.
First, due to interpolations among scales, some small-scale features may be smoothed out.
Currently, this can only be improved by increasing the sample resolution of the corresponding scale.
Second, since we use D3Q27 lattice structure, and we always overlay small scales onto large ones, more memory will be used than the corresponding INSE approaches.
Third, although we parallelize our method on the GPU, the computation is still serialized among different scales, which could be further accelerated in the future.

\section{Conclusion}

In this paper, we propose a novel continuous-scale kinetic approach to simulating fluid flows with flexible sample placement.
To significantly increase the stability and accuracy for turbulent flow simulations with kinetic approaches, we propose to employ a non-orthogonal central-moment relaxation model where we develop a novel adaptive relaxation method to retain stability while preserving sufficient turbulence details.
To respect scale continuity, we propose a new continuous-scale formulation that can easily combine different scales together with loose geometrical constraint, which directly leads to a flexible scale construction and refinement scheme to adapt scales according to the domain geometries and the flows in simulation in a more continuous manner.
The application to smoke simulations demonstrates the effectiveness of our method, with comprehensive comparisons to the existing methods to verify our advantages.



\section*{Acknowledgment}
The authors would like to thank all reviewers for their constructive comments, as well as Dr. Xinxin Zhang from University of British Columbia for sharing his fluid simulation codes for comparisons.
This work is supported by the National Natural Science Foundation of China (NSFC) - Outstanding Youth Foundation (Grant No. 61502305), as well as the startup funding of ShanghaiTech University.

\ifCLASSOPTIONcaptionsoff
  \newpage
\fi


\appendix
In our continuous-scale kinetic fluid simulation, we use the non-orthogonal central moment relaxation (CMR) model for the collision~\cite{Rosis-2017}. 
Here we give a more detailed description about the construction and application of the model.
For the derivation, please refer to the paper directly.
Like in MRT model, the CMR model first define the central moment space with the following lattice velocity definition based on the D3Q27 lattice structure: 
\begin{equation}
	\begin{aligned}
		\mathbf{c}_{x}=[0, 1, -1, &0, 0, 0, 0, 1, -1, 1, -1, 1, -1, 1, -1, \\
		&0, 0, 0, 0, 1, -1, 1, -1, 1, -1, 1, -1]^T ,\\
		\mathbf{c}_{y}=[0, 0, 0, 1, &-1, 0, 0, 1, 1,-1,-1, 0, 0, 0, 0, 1,\\
		& -1, 1, -1, 1, 1, -1, -1, 1, 1, -1, -1]^T ,\\
		\mathbf{c}_{z}=[0, 0, 0, 0, &0, 1, -1, 0, 0, 0, 0, 1, 1, 1, -1, 1, \\
		&1, -1, -1, 1, 1, 1, 1, -1, -1, -1, -1]^T ,
	\end{aligned}
\end{equation}
where $\mathbf{c}_{x}$, $\mathbf{c}_{y}$ and $\mathbf{c}_{z}$ are the x-, y-, z-component of the 27 lattice velocities $\mathbf{c}_i$.
These velocities are then translated with the flow velocity $\mathbf{u}$ to define a set of translated lattice velocities:
\begin{equation}
	\bar{\mathbf{c}}_i=\mathbf{c}_i-\mathbf{u} .
\end{equation}
The central moments are then defined by constructing a transformation matrix $\mathbf{M}$ which transforms the velocity distribution functions to central moment space as:
\begin{equation}
	\mathbf{m}_i=\mathbf{M}^T\mathbf{f} ,
\end{equation}
where $\mathbf{f}=[f_0, f_1, ..., f_{26}]^T$ and $\mathbf{m}=[\mathbf{m}_0, \mathbf{m}_1, ..., \mathbf{m}_{26}]^T$ are vectors collecting all the components of the distribution functions and their corresponding central moments, and each component of $\mathbf{M}_{ij}$ is defined as:
\begin{equation}
	\mathbf{M}_{i,j}=\bar{\mathbf{c}}^m_{x,i} \bar{\mathbf{c}}^n_{y,i}\bar{\mathbf{c}}^p_{z,i} ,
\end{equation}
where $\{x,y,z\}$ indexes the corresponding velocity component; $j=(m+1)(n+1)(p+1)-1$; and $m,n,p \in \{0,1,2\}$.
Note that $i\in{0,1,...,26}$ indexes the 27 velocities and $j=(m+1)(n+1)(p+1)-1\in{0,1,...,26}$ indexes different moments.
By expanding different orders of moments, the specific forms for $\mathbf{M}_{ij}$ is defined as follows:
\begin{equation}
	\begin{aligned}
		\mathbf{M}_{i,0} &=\mathbf{c}_{i}^0  , \quad	\mathbf{M}_{i,1} =\bar{\mathbf{c}}_{xi}  , \quad 	\mathbf{M}_{i,2} =\bar{\mathbf{c}}_{yi}  ,\\
		\mathbf{M}_{i,3} &=\bar{\mathbf{c}}_{zi}  , \quad \mathbf{M}_{i,4} =\bar{\mathbf{c}}_{xi}\bar{\mathbf{c}}_{yi}  ,\\
		\mathbf{M}_{i,5} &=\bar{\mathbf{c}}_{xi}\bar{\mathbf{c}}_{zi}  , \quad \mathbf{M}_{i,6} =\bar{\mathbf{c}}_{yi}\bar{\mathbf{c}}_{zi}  ,\\
		\mathbf{M}_{i,7} &=\bar{\mathbf{c}}_{xi}^2-\bar{\mathbf{c}}_{yi}^2  ,\quad	\mathbf{M}_{i,8} =\bar{\mathbf{c}}_{xi}^2-\bar{\mathbf{c}}_{zi}^2  ,\\
		\mathbf{M}_{i,9} &=\bar{\mathbf{c}}_{xi}^2+\bar{\mathbf{c}}_{yi}^2+\bar{\mathbf{c}}_{zi}^2  ,	\quad	\mathbf{M}_{i,10}    =\bar{\mathbf{c}}_{xi}\bar{\mathbf{c}}_{yi}^2+\bar{\mathbf{c}}_{xi}\bar{\mathbf{c}}_{zi}^2  ,\\
		\mathbf{M}_{i,11}    &=\bar{\mathbf{c}}_{xi}^2\bar{\mathbf{c}}_{yi}+\bar{\mathbf{c}}_{yi}\bar{\mathbf{c}}_{zi}^2  , \quad
		\mathbf{M}_{i,12}   =\bar{\mathbf{c}}_{xi}^2\bar{\mathbf{c}}_{zi}+\bar{\mathbf{c}}_{yi}^2\bar{\mathbf{c}}_{zi}  ,\\
		\mathbf{M}_{i,13}    &=\bar{\mathbf{c}}_{xi}\bar{\mathbf{c}}_{yi}^2-\bar{\mathbf{c}}_{xi}\bar{\mathbf{c}}_{zi}^2  ,\quad
		\mathbf{M}_{i,14}    =\bar{\mathbf{c}}_{xi}^2\bar{\mathbf{c}}_{yi}-\bar{\mathbf{c}}_{yi}\bar{\mathbf{c}}_{zi}^2  ,\\
		\mathbf{M}_{i,15}    &=\bar{\mathbf{c}}_{xi}^2\bar{\mathbf{c}}_{zi}-\bar{\mathbf{c}}_{yi}^2\bar{\mathbf{c}}_{zi}  ,\quad
		\mathbf{M}_{i,16}    =\bar{\mathbf{c}}_{xi}\bar{\mathbf{c}}_{yi}\bar{\mathbf{c}}_{zi}  ,\\
		\mathbf{M}_{i,17}    &=\bar{\mathbf{c}}_{xi}^2\bar{\mathbf{c}}_{yi}^2+\bar{\mathbf{c}}_{xi}^2\bar{\mathbf{c}}_{zi}^2
		+\bar{\mathbf{c}}_{yi}^2\bar{\mathbf{c}}_{zi}^2  ,\\
		\mathbf{M}_{i,18}    &=\bar{\mathbf{c}}_{xi}^2\bar{\mathbf{c}}_{yi}^2+\bar{\mathbf{c}}_{xi}^2\bar{\mathbf{c}}_{zi}^2
		-\bar{\mathbf{c}}_{yi}^2\bar{\mathbf{c}}_{zi}^2  ,\quad
		\mathbf{M}_{i,19}    =\bar{\mathbf{c}}_{xi}^2\bar{\mathbf{c}}_{yi}^2-\bar{\mathbf{c}}_{xi}^2\bar{\mathbf{c}}_{zi}^2  ,\\
		\mathbf{M}_{i,20}    &=\bar{\mathbf{c}}_{xi}^2\bar{\mathbf{c}}_{yi}\bar{\mathbf{c}}_{zi}^2  ,\quad
		\mathbf{M}_{i,21}    =\bar{\mathbf{c}}_{xi}\bar{\mathbf{c}}_{yi}^2\bar{\mathbf{c}}_{zi}  ,\\
		\mathbf{M}_{i,22}    &=\bar{\mathbf{c}}_{xi}\bar{\mathbf{c}}_{yi}\bar{\mathbf{c}}_{zi}^2  ,\quad
		\mathbf{M}_{i,23}    =\bar{\mathbf{c}}_{xi}\bar{\mathbf{c}}_{yi}^2\bar{\mathbf{c}}_{zi}^2  ,\\
		\mathbf{M}_{i,24}  &=\bar{\mathbf{c}}_{xi}^2\bar{\mathbf{c}}_{yi}\bar{\mathbf{c}}_{zi}^2  ,\quad
		\mathbf{M}_{i,25}    =\bar{\mathbf{c}}_{xi}^2\bar{\mathbf{c}}_{yi}^2\bar{\mathbf{c}}_{zi}  ,\\
		\mathbf{M}_{i,26}    &=\bar{\mathbf{c}}_{xi}^2\bar{\mathbf{c}}_{yi}^2\bar{\mathbf{c}}_{zi}^2  ,\\
	\end{aligned}
\end{equation}
where the vector $\mathbf{c}_{i}^0 = \mathbf{1}$.
After we convert the distribution functions $\mathbf{f}$ to the central moment space $\mathbf{m}$ by $\mathbf{m}=\mathbf{M}^T\mathbf{f}$, we can model the collision operator $\mathbf{\Omega}$ which gathers all the collision operators for each $f_i$ as:
\begin{equation}
	\mathbf{\Omega}=-\mathbf{T}\mathbf{S}(\mathbf{m}-\mathbf{m}^{eq}) ,
\end{equation}
where $\mathbf{S}$ is a diagonal matrix defining the relaxation parameters for different orders of moments and $\mathbf{m}_{i}^{eq}$ is a vector defining the equilibrium state in central moment space by:
\begin{equation}
	\begin{aligned}
		\mathbf{m}_{0}^{eq} &=\rho,  \\
		\mathbf{m}_{1}^{eq} &= 	\mathbf{m}_{2}^{eq} =	\mathbf{m}_{3}^{eq}=	\mathbf{m}_{4}^{eq}=	\mathbf{m}_{5}^{eq}=	\mathbf{m}_{6}^{eq}=	\mathbf{m}_{7}^{eq}= \mathbf{m}_{8}^{eq} =0 , \\
		\mathbf{m}_{9}^{eq} &=\rho , \\
		\mathbf{m}_{10}^{eq} &=-\rho\mathbf{u}_{x}(\mathbf{u}_{y}^2+\mathbf{u}_{z}^2) , \\
		\mathbf{m}_{11}^{eq} &=-\rho\mathbf{u}_{y}(\mathbf{u}_{x}^2+\mathbf{u}_{z}^2) , \\
		\mathbf{m}_{12}^{eq} &=-\rho\mathbf{u}_{z}(\mathbf{u}_{x}^2+\mathbf{u}_{y}^2) , \\
		\mathbf{m}_{13}^{eq} &=-\rho\mathbf{u}_{x}(\mathbf{u}_{y}^2-\mathbf{u}_{z}^2) , \\
		\mathbf{m}_{14}^{eq} &=-\rho\mathbf{u}_{y}(\mathbf{u}_{x}^2-\mathbf{u}_{z}^2) , \\
		\mathbf{m}_{15}^{eq} &=-\rho\mathbf{u}_{z}(\mathbf{u}_{x}^2-\mathbf{u}_{y}^2) , \\
		\mathbf{m}_{16}^{eq} &=-\rho\mathbf{u}_{x}\mathbf{u}_{y}\mathbf{u}_{z} , \\
		\mathbf{m}_{17}^{eq} &=\frac{\rho}{3}(9\mathbf{u}_{x}^2\mathbf{u}_{y}^2+9\mathbf{u}_{x}^2\mathbf{u}_{z}^2+9\mathbf{u}_{y}^2\mathbf{u}_{z}^2+1) , \\
		\mathbf{m}_{18}^{eq} &=\frac{\rho}{9}(27\mathbf{u}_{x}^2\mathbf{u}_{y}^2+27\mathbf{u}_{x}^2\mathbf{u}_{z}^2-27\mathbf{u}_{y}^2\mathbf{u}_{z}^2+1) , \\
		\mathbf{m}_{19}^{eq} &=3\rho\mathbf{u}_{x}^2(\mathbf{u}_{y}^2-\mathbf{u}_{z}^2) , \\
		\mathbf{m}_{20}^{eq} &=3\rho\mathbf{u}_{x}^2\mathbf{u}_{y}\mathbf{u}_{z} , \\
		\mathbf{m}_{21}^{eq} &=3\rho\mathbf{u}_{x}\mathbf{u}_{y}^2\mathbf{u}_{z} , \\
		\mathbf{m}_{22}^{eq} &=3\rho\mathbf{u}_{x}\mathbf{u}_{y}\mathbf{u}_{z}^2 , \\
		\mathbf{m}_{23}^{eq} &=-\frac{\rho}{3}\mathbf{u}_{x}(18\mathbf{u}_{y}^2\mathbf{u}_{z}^2+\mathbf{u}_{y}^2+\mathbf{u}_{z}^2) , \\
		\mathbf{m}_{24}^{eq} &=-\frac{\rho}{3}\mathbf{u}_{y}(18\mathbf{u}_{x}^2\mathbf{u}_{z}^2+\mathbf{u}_{x}^2+\mathbf{u}_{z}^2) , \\
		\mathbf{m}_{25}^{eq} &=-\frac{\rho}{3}\mathbf{u}_{z}(18\mathbf{u}_{x}^2\mathbf{u}_{y}^2+\mathbf{u}_{x}^2+\mathbf{u}_{y}^2) , \\
		\mathbf{m}_{26}^{eq} &=\rho(10\mathbf{u}_{x}^2\mathbf{u}_{y}^2\mathbf{u}_{z}^2+\mathbf{u}_{x}^2\mathbf{u}_{y}^2+\mathbf{u}_{x}^2\mathbf{u}_{z}^2+\mathbf{u}_{y}^2\mathbf{u}_{z}^2+\frac{1}{27}), \\
	\end{aligned}
\end{equation}
and $\mathbf{T}$ is an inverse transformation matrix to transform moment space vectors back to distribution functions, which is defined as $\mathbf{T}=(\mathbf{M^T})^{-1}$.
Originally, since $\mathbf{M}$ is defined with $\bar{\mathbf{c}}_i$ which is related to the macroscopic velocity $\mathbf{u}$, it is spatially and temporally varying, and $\mathbf{T}$ must be solved for every iteration, which is costly.
However, analytical expression of $\mathbf{T}$ exists which can be obtained by using Matlab function call ``simplify(...)''.
Since the expression is really long, we do not include in this appendix.
After we obtain the collision operator $\mathbf{\Omega}$, the LBE is iterated as:
\begin{equation}
	f_i(\mathbf{x}+\mathbf{c}_i\Delta t , t+\Delta t) -f_i(\mathbf{x},t) =\Omega_i(\rho,\mathbf{u}).
\end{equation}

Note that there are 27 relaxation parameters in $\mathbf{S}$ that should be further specified.
By definition, $\mathbf{m}_0$ to $\mathbf{m}_3$ correspond to macroscopic density $\rho$ and velocity $\mathbf{u}$ which are all conserved.
Thus, any values for these relaxation parameters can be used, and we set $\mathbf{S}_0$ to $\mathbf{S}_3$ to be 0.
$\mathbf{m}_4$ to $\mathbf{m}_8$ correspond to physical stress terms, which should be relaxed by macroscopic viscosity $\nu$ which is defined as
\begin{equation}
	\mathbf{S}_i = \left(3\nu+1/2\right)^{-1}, \;\;\; i\in{4,5,6,7,8}.
\end{equation}
For other components in central moment space $\mathbf{m}_9$ to $\mathbf{m}_26$, they correspond to higher order moments, their relaxation parameters $\mathbf{m}_9$ to $\mathbf{m}_{26}$ are determined by some artificial viscosities $\nu_i'$ as:
\begin{equation}
	\mathbf{S}_i = \left(3\nu_i'+1/2\right)^{-1}, \;\;\; i\in{9,10,...,26}.
\end{equation}
As argued in the main paper, the responsibility of $\nu_i'$ is to damp out higher order oscillations.
The higher the order in moment construction, the larger the artificial viscosity $\nu_i'$ should be given.
In practice, we should gradually increase $\nu_i'$ from small value to relatively large value to stabilize the dynamics by suppressing high order oscillations while maintaining sufficient accuracy.
In our method, we use the four artificial viscosities which interpolate between the lowest and highest orders in moment space, and the specific setting used in our experiments are as follows:
\begin{equation}
	\begin{aligned}
		\nu_i' &= 0.005, \;\;\; i\in{9,10,...,16}, \\
		\nu_i' &= 0.007, \;\;\; i\in{17,18,...,22}, \\
		\nu_i' &= 0.009, \;\;\; i\in{23,24,...,25}, \\
		\nu_i' &= 0.01, \;\;\;\;\; i=26,
	\end{aligned}
\end{equation}
which are a good balance between stability and accuracy, and from our various experimental results, together with the adaptive relaxation scheme, the whole dynamics is stable and accurate enough without any blow-up, even with complex geometrical boundaries.




%




\vfill


\end{document}